\documentclass[12pt,a4paper]{article}
\pdfoutput=1
\usepackage{color}
\usepackage{amssymb,amsmath,bm,bbold}
\usepackage{epsf}
\usepackage{epsfig}
\usepackage{relsize}
\usepackage[dvipsnames]{xcolor}
\usepackage[linktoc=page,bookmarks=false,colorlinks=false,linkbordercolor=RoyalBlue,citebordercolor=ForestGreen,urlbordercolor=CornflowerBlue]{hyperref}
\usepackage{latexsym,mathrsfs,dsfont}
\usepackage[normalem]{ulem} 
\usepackage[compress]{cite}
\usepackage{graphicx}
\usepackage{url}
\usepackage{booktabs}
\usepackage{float}
\usepackage{multirow}
\usepackage{changepage}
\usepackage[hypcap]{caption, subcaption}

\usepackage[titles]{tocloft}

\usepackage{tabularx,colortbl}

\usepackage{fancyhdr}
\advance \headheight by 3.0truept       

\allowdisplaybreaks[1]

\definecolor{red}{cmyk}{0,1,1,0.4}
\definecolor{darkgreen}{rgb}{0.0,0.6,0.0}
\definecolor{cDarkGrey}{RGB}{91,91,91}
\definecolor{cGrey}{RGB}{245,243,238}
\definecolor{cBlue}{RGB}{0,110,191}
\definecolor{cLightBlue}{RGB}{214,237,252}
\definecolor{cRed}{RGB}{196,0,100}
\definecolor{cLightRed}{RGB}{254,222,237}
\definecolor{cGreen}{RGB}{0,166,80}
\definecolor{cLightGreen}{RGB}{254,222,237}
\definecolor{cOrange}{RGB}{221,74,44}
\definecolor{cLightOrange}{RGB}{255,215,210}
\definecolor{cPurple}{RGB}{93,35,125}
\definecolor{cLightPurple}{RGB}{241,230,252}
\definecolor{cYellow}{RGB}{252,191,10}
\definecolor{cISSRBlue}{RGB}{0,111,174}
\definecolor{cISSRGrey}{RGB}{167,169,172}

\newcommand{\beq}{\begin{equation}}
\newcommand{\eeq}{\end{equation}}
\newcommand{\be}{\begin{equation}}
\newcommand{\ee}{\end{equation}}
\newcommand{\bi}{\begin{itemize}}
\newcommand{\ei}{\end{itemize}}
\newcommand{\ba}{\begin{array}}
\newcommand{\ea}{\end{array}}
\newcommand{\beqa}{\begin{eqnarray}}
\newcommand{\eeqa}{\end{eqnarray}}
\newcommand{\bea}{\begin{eqnarray}}
\newcommand{\eea}{\end{eqnarray}}
\newcommand{\beqn}{\begin{eqnarray}}
\newcommand{\eeqn}{\end{eqnarray}}

\newcommand{\nn}{\nonumber}

\newcounter{TODO}

\newcommand{\ord}{{\cal O}}

\newcommand{\GeV}{\,\text{GeV}}

\newcommand{\teV}{\text{TeV}}

\newcommand{\vcb}{|V_{cb}|}
\newcommand{\vtd}{|V_{td}|}
\newcommand{\vub}{|V_{ub}|}
\newcommand{\vts}{|V_{ts}|}
\newcommand{\vus}{|V_{us}|}
\newcommand{\vud}{|V_{ud}|}
\newcommand{\vcd}{|V_{cd}|}

\newcommand{\bsi}{B_6^{(1/2)}}
\newcommand{\bei}{B_8^{(3/2)}}
\newcommand{\kepe}{\kappa_{\varepsilon^\prime}}

\newcommand{\epe}{\varepsilon'/\varepsilon}

\def\kpn{K^+\rightarrow\pi^+\nu\bar\nu}
\def\klpn{K_{L}\rightarrow\pi^0\nu\bar\nu}

\newcommand{\IM}{\rm{Im}}
\newcommand{\RE}{\rm{Re}}

\newcommand{\hatV}{{\hat V}}

\begin{document}

\begin{flushleft}
  {\em Version of \today}
\end{flushleft}

\vspace{-14mm}
\begin{flushright}
  AJB-23-6
\end{flushright}

\medskip

\begin{center}
{\Large\bf\boldmath
  Kaon Theory: 50 Years Later*
}
\\[1.0cm]
{\bf \large
    Andrzej~J.~Buras
}\\[0.3cm]

{\small
TUM Institute for Advanced Study,
    Lichtenbergstr. 2a, D-85748 Garching, Germany \\[0.2cm]
Physik Department, TUM School of Natural Sciences, TU M\"unchen,\\ James-Franck-Stra{\ss}e, D-85748 Garching, Germany
}

\end{center}

\vskip 0.5cm

\begin{abstract}
  \noindent
  We summarize the status of the Kaon Theory 50 years after the seminal paper
  of Kobayashi and Maskawa who pointed out that six quarks are necessary
  to have CP violation in the Standard Model (SM) and presented a parametrization   of a $3\times 3$ unitary matrix that after the discovery of the charm
  quark in 1974 and the $b$ quark in 1977 dominated the field of flavour
  changing processes. One of the main goals of flavour physics since then
  was  the   determination of the four parameters of this matrix, which we will
  choose here to be $|V_{us}|$, $|V_{cb}|$ and the two angles of the unitarity triangle, $\beta$ and $\gamma$ with $\vus$ introduced by Cabibbo in 1963.
  I will  summarize recent strategy for determination of these parameters without new physics (NP) infection. It is based on  the conjecture of the
  absence of relevant NP contributions to $\Delta F=2$ processes that   indeed can be demonstrated by a negative rapid test: the
  $|V_{cb}|-\gamma$ plot.
  This in turn allows  to obtain SM predictions for   rare $K$ and $B$ decays that are most precise to date. We present strategies for the explanation of the  anticipated   anomaly in the ratio $\varepsilon'/\varepsilon$ and 
  the observed    anomalies in $b\to s\mu^+\mu^-$ transitions that are consistent with our  $\Delta F=2$ conjecture. In particular, the absence
  of NP in the parameter $\varepsilon_K$, still allows for significant NP
  effects in $\varepsilon'/\varepsilon$ and in rare Kaon decays,
  moreover in a correlated manner. Similar the absence
  of NP in $\Delta M_s$ combined with  anomalies in $b\to s\mu^+\mu^-$ transitions hints for the presence of right-handed quark currents.
    We also discuss how the nature
  of neutrinos, Dirac vs. Majorana one, can be probed in $K\to\pi\nu\bar\nu$
  and $B\to K(K^*)\nu\bar\nu$ decays. The present status of
  the $\Delta I=1/2$ rule and of $\varepsilon'/\varepsilon$ is summarized.
\end{abstract}

*) Talk given at the KM50 Symposium, KEK, Tsukuba, February 11th, 2023

\thispagestyle{empty}
\newpage
\setcounter{page}{1}
\section{Overture}
Our story begins in 1963 when Cabibbo  \cite{Cabibbo:1963yz} introduced the angle $\theta_C$ which could be determined in tree-level Kaon decays. This 
allowed to estimate roughly the $K_L-K_S$ mass difference $\Delta M_K$ and also the branching ratio for the $K_L\to\mu^+\mu^-$ decay. With 
only the existence of three quarks, $u,d,s$,  known in the 1960s, these
estimates turned out to be significantly higher than the data. These
days we would conclude that some new physics (NP) is required to suppress
these observables to agree with the experimental data. This NP turned out to be the charm quark. Indeed, this problem has been
solved by Glashow, Iliopoulos and Maiani (GIM) \cite{Glashow:1970gm} who, adding the charm quark and requiring the resulting $2\times 2$  matrix to be unitary, outlined how  the two observables could be suppressed.
But only in 1974 Gaillard and Lee \cite{Gaillard:1974hs} presented explicit calculations of both observables in the four-quark model  and with the help of the $\Delta M_K$ could  predict the charm quark mass prior to its discovery.
In particular  $\Delta M_K$ was found to be suppressed by the ratio $m_c^2/M_W^2$ in agreement with its experimental value.

However, already in February 1973, Kobayashi and Maskawa pointed out in a seminal
paper \cite{Kobayashi:1973fv} that with four quarks CP-violation in $K_L\to\pi\pi$ decays, discovered in 1964 \cite{Christenson:1964fg}, could not be explained and that additional NP
was required to do it: two additional quarks known these days under the names
of the beauty quark and the top quark that were discovered in 1977 and 1995, respectively.
In this context they generalized the unitary $2\times 2$ matrix used in the
GIM paper to a unitary $3\times 3$ matrix that guaranteed the absence of
FCNC processes at the tree-level and also their suppression at the loop level
provided the masses of exchanged quarks were significantly smaller than $M_W$.
To my knowledge the pioneering phenomenological analysis of the KM scenario
has been performed by Ellis, Gaillard and Nanopoulos \cite{Ellis:1976fn} in 1976, in fact during
my CERN fellow days. Because the top quark mass was chosen to be below $20\GeV$ in this paper and the parameters of the CKM matrix  \cite{Cabibbo:1963yz,Kobayashi:1973fv}  were very weakly constrained, this analysis does not resemble similar analyses performed already in the 1980s after it has been found that
$\vcb$ is much smaller than $\vus$, and in particular after the top quark discovery in 1995. But this was the first analysis of this type.

Before I move to the year 2023 let me mention still the  papers which 
 certainly  played an important role in the phenomenological flavour analyses already for many years
\begin{itemize}
\item
  The standard parametrization of the CKM matrix of Chau and Keung \cite{Chau:1984fp}.
\item
  Its very approximated form of Wolfenstein  \cite{Wolfenstein:1983yz} in which
  the parameters $\lambda$, $A$, $\varrho$ and  $\eta$ have been introduced.
\item
  Important papers by Jarlskog on the invariant measure of CP violation
  in the SM
  \cite{Jarlskog:1985ht,Jarlskog:1985cw}.
\item
 First papers on Unitarity Triangles \cite{Jarlskog:1988ii,Aleksan:1994if}.
\item
  Much more precise Wolfenstein-like parametrization introduced in Munich in 1994 \cite{Buras:1994ec} in which in particular the apex of the unitarity triangle
  is described by $(\bar\varrho,\bar\eta)$. A similar parametrization has been proposed by Branco and Lavoura \cite{Branco:1988ba}.
\item
  The concept of Minimal Flavour Violation
  \cite{Chivukula:1987py,Buras:2000dm,DAmbrosio:2002vsn,Kagan:2009bn}
    which implies strict correlations between flavour changing processes in different meson systems.
 \end{itemize}
All these topics and many other are discussed in detail in my recent book \cite{Buras:2020xsm}. See also other books \cite{Branco:1999fs,Bigi:2000yz,Bigi:2021hxw}.

Kaon physics is not as rich as $B$ physics but still has a number of stars
that have been considered by experimentalists and theorists in the last four decades. These
are
\be
\kpn,\qquad \klpn, \qquad K_{L,S}\to \mu^+\mu^-,\qquad K_L\to\pi^0\ell^+\ell^-,
\ee
and
\be
\Delta M_K, \qquad \varepsilon_K,\qquad \epe, \qquad \Delta I=1/2~\text{Rule}\,.
\ee

They all can give important information about very short distance scales but
to identify NP, correlations with $B_{s,d}$, $B^+$ and $D$ observables, electric dipole moments and lepton physics are crucial \cite{Buras:2020xsm,Buras:2013ooa}.
The fortunate side of this field, with the exception of the ratio $\epe$ and
the $\Delta I=1/2$ rule, is the theoretical cleanness of these decays.
On the other hand, the unfortunate side of it are very low branching ratios of the decays involved. As an example I am waiting already 30 years
to be able to compare my results with experiment for a number of
rare $K$ and $B$ decay branching ratios\footnote{My interest in rare Kaon decays increased by much during the 1988 Kaon conference in Vancouver, in particular after the talks of Larry Littenberg on $\klpn$ and of Fred Gilman on $K_L\to\pi^0\ell^+\ell^-$.}. This is the case of  the first  papers 
on the next-to-leading order (NLO) strong interaction (QCD) corrections to rare decay branching ratios for $\kpn$, $\klpn$,
$B_{s,d}\to \mu^+\mu^-$ and alike calculated in collaboration  with Gerhard Buchalla in 1993 \cite{Buchalla:1992zm,Buchalla:1993bv} and  numerous papers with him and other collaborators in the last 30 years including
two reviews in 2004 and 2022 \cite{Buras:2004uu,Aebischer:2022vky}.
It is then evident that QCD plays a very important role in this field. It happens that also QCD celebrates its 50th birthday which is documented in \cite{Gross:2022hyw}.

There are a number of reviews that touch on the topics discussed here. This is
in particular \cite{Goudzovski:2022scl} and the talk by Taku Yamanaka at this symposium. On the other hand the search for feebly-interacting light particles like axions, axinos and dark matter are beyond the scope of our review. A nice description of this topic
can be found in \cite{Lanfranchi:2020crw,Goudzovski:2022vbt}.

The overture is finished and we move to the KM symphony.

\begin{table}[thb]
\begin{center}
\begin{tabular}{|l|l|l|}
\hline
\bf \phantom{XXXX} Decay &  {\bf NLO} & {\bf NNLO}  \\
\hline
\hline
Current-Current $(Q_1,Q_2)$      &\cite{Altarelli:1980te,Buras:1989xd}& \cite{Gorbahn:2004my}\\
QCD penguins $(Q_3,Q_4,Q_5,Q_6)$  &\cite{Buras:1991jm,Buras:1992tc,Buras:1993dy,Ciuchini:1992tj,Ciuchini:1993vr,Chetyrkin:1997gb},\cite{Fleischer:1992gp}& \cite{Gorbahn:2004my,Cerda-Sevilla:2016yzo}\\
electroweak penguins $(Q_7,Q_8,Q_9,Q_{10})$  &\cite{Buras:1992zv,Buras:1993dy,Ciuchini:1992tj,Ciuchini:1993vr} & \cite{Buras:1999st}\\
inclusive non-leptonic decays       & \cite{Altarelli:1980te,Altarelli:1980fi,Buchalla:1992gc,Bagan:1994zd,Bagan:1995yf,Krinner:2013cja}; \cite{Jamin:1994sv}  & \\
Current-Current (BSM) & \cite{Ciuchini:1997bw,Buras:2000if}    & \\
Penguins (BSM)  &\cite{Buras:2000if}  &\\
\hline
\end{tabular}
\end{center}
\caption{NLO and NNLO Calculations for Non-leptonic and Semi-Leptonic $\Delta F=1$ Transitions. References on semi-leptonic $B$ decays can be found in my
    book \cite{Buras:2020xsm} and in the recent review \cite{Buras:2011we}.\label{TAB1}}
\end{table}

\begin{table}[thb]
\begin{center}
\begin{tabular}{|l|l|l|l|}
\hline
\bf \phantom{XXXXX} Decay &  {\bf NLO QCD} & {\bf NNLO QCD} & {\bf EW} \\
\hline
\hline
$\eta_1$                   & \cite{Herrlich:1993yv}&  \cite{Brod:2011ty}&\\
$\eta_2,~\eta_B$           & \cite{Buras:1990fn} & & \cite{Gambino:1998rt,Brod:2021qvc}\\
$\eta_3$                   & \cite{Herrlich:1995hh,Herrlich:1996vf}& 
\cite{Brod:2010mj}& \cite{Brod:2022har}\\
$\eta_{tt}$         & \cite{Brod:2019rzc}      & & \cite{Brod:2021qvc}  \\
$\eta_{ut}$         &   \cite{Brod:2019rzc}     & \cite{Brod:2019rzc} & \cite{Brod:2022har} \\
ADMs BSM & \cite{Ciuchini:1997bw,Buras:2000if}  & & \\
\hline
$\Delta\Gamma_{B_s}$      & 
\cite{Beneke:1998sy,Ciuchini:2003ww,Beneke:2003az,Lenz:2006hd,Asatrian:2017qaz,Asatrian:2020zxa,Gerlach:2021xtb,Gerlach:2022wgb}&\cite{Gerlach:2022hoj} & \\
$\Delta\Gamma_{B_d}$      & \cite{Ciuchini:2003ww,Beneke:2003az}&& \\
Lifetime Ratios &  \cite{Keum:1998fd,Ciuchini:2001vx,Beneke:2002rj,Franco:2002fc,Kirk:2017juj,King:2021jsq} &&\\
$\Delta F=2$ Tree-Level & \cite{Buras:2012fs} & &  \\
\hline
\end{tabular}
\end{center}
\caption{NLO and NNLO Calculations for $\Delta F=2$ and $\Delta F=0$
  Transitions.\label{TAB2}}
\end{table}

\begin{table}[thb]
\begin{center}
\begin{tabular}{|l|l|l|}
\hline
\bf \phantom{XXXXX} Decay &  {\bf NLO} & {\bf NNLO}  \\
\hline
\hline
$K^0_L \rightarrow \pi^0\nu\bar{\nu}$, 
$B \rightarrow X_{\rm s}\nu\bar{\nu}$  &\cite{Buchalla:1992zm,Buchalla:1993bv,Misiak:1999yg,Buchalla:1998ba}& \\
$K^+   \rightarrow \pi^+\nu\bar{\nu}$  & \cite{Buchalla:1993wq,Buchalla:1998ba} & \cite{Buras:2006gb}\\
$K_{\rm L} \rightarrow \pi^0\ell^+\ell^-$  & \cite{Buras:1994qa} & \\
$B_{s,d} \rightarrow l^+l^-$  &\cite{Buchalla:1992zm,Buchalla:1993bv,Misiak:1999yg,Buchalla:1998ba}& \cite{Hermann:2013kca}\\
 $K_{\rm L} \rightarrow \mu^+\mu^-$ & \cite{Buchalla:1993wq,Buchalla:1998ba} &
\cite{Gorbahn:2006bm}\\
$K^+\to\pi^+\mu\bar\mu$               & \cite{Buchalla:1994ix} & \\
EW to Charm in $K^+   \rightarrow \pi^+\nu\bar{\nu}$ & \cite{Brod:2008ss}  &    \\
EW to Top in $K\to\pi\nu\bar\nu$ & \cite{Buchalla:1997kz,Brod:2010hi}  &  \\
EW to Top in $B_{s,d} \rightarrow l^+l^-$ & 
\cite{Buchalla:1997kz,Bobeth:2013tba}  &  \\
\hline
\end{tabular}
\end{center}
\caption{NLO and NNLO Calculations for Rare K and B decays. \label{TAB3}}
\end{table}

\section{Kobayashi-Maskawa Symphony}
\subsection{Theoretical Framework}\label{THF}
This framework is based on the operator product expansion \cite{Wilson:1969zs,Wilson:1972ee,Zimmermann:1972tv} that allows to separate
the calculation of various flavour observables like decay branching ratios into short distance ones represented
by the  Wilson coefficients of the involved local operators and the long distance ones contained in the hadronic matrix elements of these operators. 
I will be very brief about this topic because it is rather technical and is
well documented in the literature. I do it despite the fact that I spent  more than a decade calculating NLO QCD corrections to  the Wilson coefficients relevant  for quark mixing and  rare  $K$ and $B$ decays, not
only in the SM but also beyond it. The first extensive review of this
topic has been presented by Buchalla, Lautenbacher and myself in 1995 in  \cite{Buchalla:1995vs} and more pedagogically in  my Les Houches lectures in 1998  \cite{Buras:1998raa}. More up to date summary
can be found in my book \cite{Buras:2020xsm} and in particular in the
most recent
review with many anecdotes that has been just  published in Physics Reports
\cite{Buras:2011we}. I would claim that presently the status of these
corrections within the SM is very good and not too bad beyond the SM.
All relevant references can be found in \cite{Buras:2011we}
but it is appropriate to collect in Tables~\ref{TAB1}, \ref{TAB2} and \ref{TAB3} those that fit best to this review.

Important progress has also been  done for  the extraction of short distance
contribution to $K_S\to\mu^+\mu^-$
\cite{DAmbrosio:2017klp,Dery:2021mct,Brod:2022khx} offering still another
precision observable in addition to $\kpn$ and $\klpn$ decays.

While these calculations improved considerably the precision of
theoretical predictions in weak decays and can be considered as an
important progress in this field, the pioneering LO QCD calculations
for current-current operators \cite{Altarelli:1974exa,Gaillard:1974nj}, penguin operators 
\cite{Shifman:1976ge,Gilman:1979bc},
$\Delta S=2$ operators \cite{Vysotsky:1979tu,Gilman:1982ap} should not be forgotten. This also is the case of LO QCD calculations of rare $K$ decays 
\cite{Dib:1988md,Flynn:1988ve,Dib:1988js,Dib:1989cc} and of the ratio $\epe$ \cite{Flynn:1989iu,Buchalla:1989we} for large $m_t$.

This takes care of short distance contributions. But
also hadronic matrix elements, in particular the ones relevant for quark mixing
are in a good shape. I refer to FLAG reports \cite{FlavourLatticeAveragingGroupFLAG:2021npn} and in particular to the article of Aida El-Khadra in this volume, where the references to the rich literature can be found. Some references will be given in the context of our presentation.

The final ingredients necessary to make predictions for $K$ meson physics and
also $B$ physics are the CKM parameters that regularly are obtained from
global fits performed by UTfitter \cite{UTfit:2022hsi}, CKMfitter  \cite{Charles:2004jd} and PDG
\cite{Workman:2022ynf}. My strategy for finding the CKM parameters differs
presently from the ones of these groups and this is the topic of the second movement.

\subsection{SM Predictions for Rare $K$ and $B$ Decays without NP \\ Infection}
\subsubsection{Problems with Present Global Fits}
In my view there are presently the following problems with global fits just listed
\cite{Buras:2022qip}:
\begin{itemize}
\item
  In a global fit which contains processes that are infected by NP the resulting
  CKM parameters are also infected by it and consequently the resulting
  branching ratios cannot be considered as genuine SM predictions.
  Consequently the resulting deviations from the SM predictions obtained in
  this manner (the pulls) are not the deviations one would find if
  the CKM parameters were not infected by NP.
\item
  Tensions in the determinations of $\vcb$ and $\vub$ from inclusive and exclusive tree-level decays \cite{Bordone:2021oof,FlavourLatticeAveragingGroupFLAG:2021npn}. Using these results lowers the precision with which CKM parameters can be
  determined and their inclusion in the fit should be avoided until theorists
  agree what the values of  $\vcb$ and $\vub$ are. This is also the view of
  a number of theorists who attempt to determine these parameters through tree-level exclusive and inclusive tree-level decays, although they did not state it in print.
\item
  Hadronic uncertainties in some observables included in the fit are much larger than in many rare $K$ and $B$ decays. Even if they can be given a lower weight
  in the fit, they lower the precision and should be presently avoided.
\end{itemize}
\subsubsection{Strategy}
In what follows I want to summarize the strategy developed in two papers
with Elena Venturini \cite{Buras:2021nns,Buras:2022wpw} which generalized my
2003 strategy used for $B_{s,d}\to\mu^+\mu^-$ decays \cite{Buras:2003td}
to all $K$ and $B$ decays. This strategy deals with the second and the third item
above but as I realized in \cite{Buras:2022qip} it  solves also the first
problem. It consists of four steps.

{\bf Step 1}

Remove CKM dependence by calculating suitable ratios of decay branching ratios  to
the mass differences $\Delta M_s$ and $\Delta M_d$ in the case of $B_s$ and
$B_d$ decays, respectively and to the parameter $|\varepsilon_K|$ in the case of Kaon decays. By suitable we mean for instance that in order to eliminate the $\vcb$
dependences in the branching ratios for $\kpn$ and $\klpn$, the parameter $|\varepsilon_K|$ 
has to be raised, as given later, to the power $0.82$ and $1.18$, respectively.
For $B_{s,d}$ decays one just divides the branching ratios by $\Delta M_{s,d}$, respectively.
In this manner CKM
dependence can be fully eliminated for all $B$ decay branching ratios. For
$K$ decays  only the dependence on the angle $\beta$ in the UT remains.
The dependence on $\gamma$ is practically absent so that future improvements
on the measurements of $\gamma$ by LHCb and Belle II collaborations will not
have any impact on these particular ratios. On the other hand improved measurements of $\beta$ and improved values of hadronic parameters will reduce
the uncertainties in these ratios. It should be emphasized that already these
ratios constitute very good tests of the SM.

{\bf Step 2}

Set $\Delta M_s$, $\Delta M_d$, $|\varepsilon_K|$ and the mixing induced
CP asymmetries  $S_{\psi K_S}$ and $S_{\psi \phi}$ to their experimental values.
This is done usually in global fits as well but here we confine the
fit to these observables. 
The justification for this step is the fact that all these observables
can be simultaneously described within the SM without
any need for NP contributions and the theoretical and experimental status
of these $\Delta F=2$ observables is exeptionally good. In turn
this step not only avoids the tensions in the determinations of $\vcb$
and $\vub$ in tree level decays, but also provides SM predictions for numerous rare $K$ and $B$  branching ratios that are most accurate to date \cite{Buras:2022wpw,Buras:2021nns,Buras:2022qip}.

{\bf Step 3}

In order to be sure  that the $\Delta F=2$ archipelago is not infected
by NP a rapid test has to be performed with the help of the $\vcb-\gamma$ plot
\cite{Buras:2022wpw,Buras:2021nns}. This test turns out to be negative
dominantly thanks to the 2+1+1 HPQCD lattice  calculations of $B_{s,d}-\bar B_{s,d}$ hadronic matrix elements \cite{Dowdall:2019bea}\footnote{Similar results for $\Delta M_d$ and $\Delta M_s$ hadronic
    matrix elements have been obtained within the HQET sum rules in
    \cite{Kirk:2017juj} and \cite{King:2019lal,King:2019rvk}, respectively.}. The superiority of the
$\vub-\gamma$ plots over UT plots in this context has been emphasized in
\cite{Buras:2022nfn}. We will present these issues in more details soon.

{\bf Step 4}  

As the previous step has lead to a negative rapid test we can now determine the
CKM parameters without NP infection on the basis of $\Delta F=2$ observables
alone. It should be noted that this step can be considered as a reduced
global fit of CKM parameters in which only $\Delta F=2$ observables have been taken into account.

All the problems listed above are avoided in this manner and having CKM
parameters at hand one can make rather precise SM predictions for the
observables outside  the $\Delta F=2$ archipelago and compare them with the
experimental data. The pulls obtained in this manner are more reliable
than the ones obtained from global fits that include several additional
observables.

In this context let us recall  the values of $\vcb$
extracted from inclusive and exclusive tree-level semi-leptonic $b\to c$ decays
\cite{Bordone:2021oof,FlavourLatticeAveragingGroupFLAG:2021npn}
\be
\vcb_\text{incl}=(42.16\pm0.50)\cdot 10^{-3},\qquad \vcb_\text{excl}=(39.21\pm0.62)\cdot 10^{-3}\,.
\ee
As FCNC processes are sensitive functions of $\vcb$, varying it from $39\cdot 10^{-3}$ to $42\cdot 10^{-3}$ changes $\Delta M_{s,d}$ and $B$-decay branching ratios
by roughly $16\%$, $\kpn$ branching ratio by $23\%$, $\varepsilon_K$ by $29\%$
and $\klpn$ and $K_S\to\mu^+\mu^-$ branching ratios by $35\%$.

These uncertainties are clearly a disaster for those like me, my collaborators
and other experts in NLO and NNLO calculations who spent decades to reduce theoretical uncertainties in basically all important rare $K$ and $B$ decays and
quark mixing observables down to $(1-2)\%$. It is also a disaster for lattice QCD experts who for quark mixing observables and in particular  meson weak decay constants achieved the accuracy at the level of a few percents. See
the article of Aida El-Khadra in this volume.

\begin{figure}
\centering
\includegraphics[width = 0.55\textwidth]{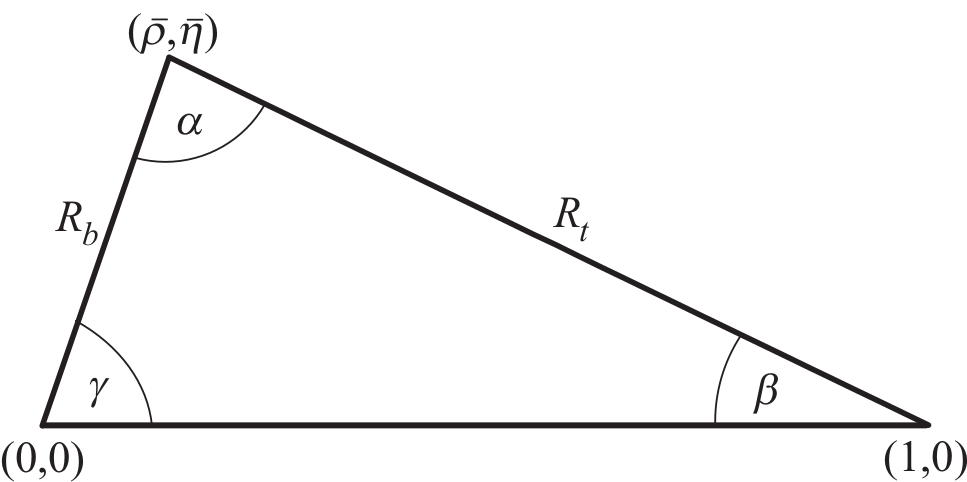}
 \caption{\it The Unitarity Triangle. }\label{UUTa}
\end{figure}

\boldmath
\subsubsection{New Formulae for $\bar\varrho$ and $\bar\eta$}
\unboldmath
Before describing the results of our strategy let us emphasize that it uses
as basic CKM parameters \cite{Blanke:2018cya,Buras:2022wpw,Buras:2021nns,Buras:2022qip}
\be\label{best}
\vus=\lambda, \qquad \vcb,\qquad  \beta,\qquad \gamma,
\ee
with $\beta$ and $\gamma$ being the two angels of the UT shown in Fig.~\ref{UUTa}.

It should be noted that all parametrizations listed at the begining of this writing involve three real parameters and one complex phase.
In particular in 1986 Harari and Leurer \cite{Harari:1986xf} recomended
the standard parametrization because of the 
relations\footnote{$c_{13}=1$ to an excellent accuracy.}
\be\label{LH}
s_{12}=\vus, \qquad   s_{13}=\vub,\qquad s_{23}=\vcb, \qquad \gamma,
\ee
 that allow the determination of  these  four parameters separately in tree-level decays.
Consequently,  basically all flavour phenomenology in the last three decades  used such sets of parameters. In particular the determination of the UT
  was dominated by the measurements of its sides $R_b$ and $R_t$ through
  tree-level $B$ decays and the $\Delta M_d/\Delta M_s$ ratio, respectively, with   some participation of the measurements of the angle $\beta$ through the mixing  induced CP-asymmetries like $S_{\psi K_S}$, the parameter $\varepsilon_K$  and much less precise angle $\gamma$. This is the case not only
  of global analyses by UTfitter\cite{Bona:2007vi} and CKMfitter \cite{Charles:2004jd}  but also of less sophisticated
  determinations of the CKM matrix and of the UT.

  However, as pointed out in  \cite{Buras:2022wpw,Buras:2021nns,Buras:2022qip}, the most powerful strategy appears eventually to be the one which
  uses as basic CKM parameters the ones in (\ref{best}),
that is two mixing angles and two phases.
This choice is superior to the one in which $\beta$ is replaced by $\vub$ for
several reasons:
\begin{itemize}
\item
  The known tensions between exclusive and inclusive determinations of $\vcb$
  and $\vub$ \cite{Bordone:2021oof,FlavourLatticeAveragingGroupFLAG:2021npn}
  are represented only by $\vcb$ which can be eliminated efficiently
  by constructing suitable ratios of flavour observables $R_i(\beta,\gamma)$, see below,   which are free of the tensions in question.
\item
  As pointed out already in 2002 \cite{Buras:2002yj},
  the most
  efficient strategy for a precise determination of the apex of the UT, that is
  $(\bar\varrho,\bar\eta)$, is to use the measurements of the angles $\beta$ and $\gamma$. Indeed, among any pairs of two variables representing the sides and the angles of the UT that are assumed for this exercise to be known with the same precision, the measurement of $(\beta,\gamma)$ results in the most accurate values of $(\bar\varrho,\bar\eta)$. 
  The second best strategy would be the measurements of $R_b$ and $\gamma$. However, in view of the tensions between different determinations of $\vub$ and  $\vcb$, that enter $R_b$,  the $(\beta,\gamma)$ strategy
will  be a clear winner once LHCb and Belle II collaborations will improve  the measurements of these two angles.
  \item
  The $\vcb-\gamma$ plots for fixed $\beta$, proposed in  \cite{Buras:2022wpw,Buras:2021nns} are, as emphasized in \cite{Buras:2022nfn}, useful companions to
  common unitarity triangle fits because they exhibit better possible inconsistences between $\vcb$ and $(\beta,\gamma)$ determinations than the latter fits.
  We will demonstrate this below.
\end{itemize}

In this context let us present two simple formulae that are central in the
$(\beta,\gamma)$ strategy as they allow to calculate the appex of the UT in no
time, but 
to my knowledge have not been presented in the literature before,
not even in our 2002 paper \cite{Buras:2002yj}. They read
\be\label{AJB23}
\boxed{\bar\varrho=\frac{\sin\beta\cos\gamma}{\sin(\beta+\gamma)},\qquad
  \bar\eta=\frac{\sin\beta\sin\gamma}{\sin(\beta+\gamma)}.}
\ee
They  follow simply from
\be\label{rb}
\bar\varrho=R_b\cos\gamma,\qquad \bar\eta=R_b\sin\gamma, \qquad
R_b=\frac{\sin\beta}{\sin(\beta+\gamma)}
\ee
with the first two relations representing $(R_b,\gamma)$ strategy \cite{Buras:2002yj}. The expression for $R_b$ has been presented already in \cite{Branco:1999fs} and possibly in other articles. 
Evidently the formulae in (\ref{AJB23}) can be derived by high-school students\footnote{In fact I recall that I had to solve such a triangle problem in 1962.}, but the UT is unknown to them and somehow no flavour physicist got the idea to present them in print so far.

Several other useful relations can be found in \cite{Buras:2002yj,Branco:1999fs}. In particular the one of $(R_t,\beta)$ strategy
\be\label{rt}
\bar\varrho=1-R_t\cos\beta,\qquad \bar\eta=R_t\sin\beta,\qquad R_t=\frac{\sin\gamma}{\sin(\beta+\gamma)}.
\ee

It should be also realized that 
  in the coming years through the precise measurements of both angles by the LHCb and Belle II collaborations the  simple formulae in (\ref{AJB23}) should
  be very useful for the construction of the UT. Also $R_b$ and $R_t$ will
  be easily found using the expressions in (\ref{rb}) and (\ref{rt}).

In Tables~\ref{tab:1} and \ref{tab:2} we show $\bar\varrho$ and $\bar\eta$ as functions of $\gamma$ for different values of $\beta$ in the expected ranges
for the latter parameters\footnote{I would like to thank Mohamed Zied Jaber for checking these tables.}. The present experimental determinations of $\beta$
and $\gamma$ read
\be\label{betagamma}
\boxed{\beta=(22.2\pm 0.7)^\circ, \qquad  \gamma = (63.8^{+3.5}_{-3.7})^\circ \,.} \ee
 Here the value for $\gamma$ is the most recent one from the LHCb which updates
the one in \cite{LHCb:2021dcr} $(65.4^{+3.8}_{-4.2})^\circ$. It is not as precise
as the one in (\ref{CKMoutput}) that follows from our strategy  but fully consistent with it. In the coming
years LHCb and Belle II should reduce the error in $\gamma$ in the ballpark
of $1^\circ$ and also the error on $\beta$ will be reduced. In fact for the
strategies of  \cite{Buras:2022wpw,Buras:2021nns,Buras:2022qip} the reduction
of the error on $\beta$ is even more important than the one on $\gamma$ because
$\beta$ is an input but $\gamma$ an output.

\begin{table}[!ht]
    \centering
    \begin{tabular}{|l|l|l|l|l|l|l|l|l|}
    \hline
        $\gamma/\beta$ & $21.6^\circ$ & $21.8^\circ$ & $22.0^\circ$ & $22.2^\circ$ &
$22.4^\circ$ & $22.6^\circ$ &$22.8^\circ$ & $23.0^\circ$\\ \hline
        $ 60^\circ$ & 0.186 & 0.188 & 0.189 & 0.191 & 0.192 & 0.194 & 0.195 & 0.197 \\
        $ 61^\circ$ & 0.180 & 0.181 & 0.183 & 0.184 & 0.186 & 0.187 & 0.189 & 0.190 \\
        $ 62^\circ$ & 0.174 & 0.175 & 0.177 & 0.178 & 0.180 & 0.181 & 0.183 & 0.184 \\
        $ 63^\circ$ & 0.168 & 0.169 & 0.171 & 0.172 & 0.174 & 0.175 & 0.176 & 0.178 \\
        $ 64^\circ$ & 0.162 & 0.163 & 0.165 & 0.166 & 0.167 & 0.169 & 0.170 & 0.172 \\
        $ 65^\circ$ & 0.156 & 0.157 & 0.159 & 0.160 & 0.161 & 0.163 & 0.164 & 0.165 \\
        $ 66^\circ$ & 0.150 & 0.151 & 0.152 & 0.154 & 0.155 & 0.156 & 0.158 & 0.159 \\
        $ 67^\circ$ & 0.144 & 0.145 & 0.146 & 0.148 & 0.149 & 0.150 & 0.151 & 0.153 \\
        $ 68^\circ$ & 0.138 & 0.139 & 0.140 & 0.142 & 0.143 & 0.144 & 0.145 & 0.146 \\ \hline
    \end{tabular}
    \caption{\it Values of $\bar\varrho$ for different values of $\beta$ and $\gamma$ .}
    \label{tab:1}
\end{table}

\begin{table}[!ht]
    \centering
    \begin{tabular}{|l|l|l|l|l|l|l|l|l|}
    \hline
       $\gamma/\beta$ & $21.6^\circ$ & $21.8^\circ$ & $22.0^\circ$ & $22.2^\circ$ &
$22.4^\circ$ & $22.6^\circ$ &$22.8^\circ$ & $23.0^\circ$\\
        \hline
         $ 60^\circ$ &$ 0.322$ &$ 0.325$ &$ 0.328 $&$ 0.330 $&$ 0.333 $&$ 0.336 $&$ 0.338$ &$ 0.340$\\
         $ 61^\circ$ & 0.325 & 0.327 & 0.330 & 0.333 & 0.336 & 0.338 & 0.341 & 0.344 \\
         $ 62^\circ$ & 0.327 & 0.330 & 0.333 & 0.335 & 0.338 & 0.341 & 0.344 & 0.346 \\
         $ 63^\circ$ & 0.329 & 0.332 & 0.335 & 0.338 & 0.341 & 0.343 & 0.346 & 0.349 \\
         $ 64^\circ$ & 0.332 & 0.335 & 0.338 & 0.340 & 0.343 & 0.346 & 0.349 & 0.352 \\
         $ 65^\circ$ & 0.334 & 0.337 & 0.340 & 0.343 & 0.346 & 0.349 & 0.351 & 0.354 \\
         $ 66^\circ$ & 0.337 & 0.340 & 0.342 & 0.345 & 0.348 & 0.351 & 0.354 & 0.357 \\
         $ 67^\circ$ & 0.339 & 0.342 & 0.345 & 0.348 & 0.351 & 0.354 & 0.357 & 0.360 \\
         $ 68^\circ$ & 0.341 & 0.344 & 0.347 & 0.350 & 0.353 & 0.356 & 0.359 & 0.362 \\ \hline
    \end{tabular}
    \caption{\it Values of $\bar\eta$ for different values of $\beta$ and $\gamma$ .}
    \label{tab:2}
\end{table}

It is interesting to compare these tables with the most recent ``angle-only fit'' of UTfitter \cite{UTfit:2022hsi}
\be\label{CKMoutput5}
{\bar\varrho=0.156(17),\qquad \bar\eta=0.341(12)\,.}
\ee
The significant error on $\bar\varrho$ will be reduced by much with the improved measurement of $\gamma$. Let us then anticipate  future measurements with
\be\label{CKMoutput26}
\gamma=64.0(10)^\circ, \qquad \beta=22.2(4)^\circ,\qquad (2026)\,.
\ee
Using (\ref{AJB}) and adding the errors in quadrature one would find
\be
\bar\varrho=0.166(7),\qquad \bar\eta=0.340(6)\,,\qquad (2026)\,.
\ee
Once the tensions in the determination of $\vcb$ and $\vub$ will be resolved
the full fit will of course result in even more precise values.

\subsubsection{Results}
 One constructs then a multitude of $\vcb$-independent ratios $R_i$ not only
of branching ratios to quark mixing observables but also of branching
ratios themselves. Those which involve branching ratios from different meson
systems depend generally on $\beta$ and $\gamma$. Once $\beta$ and $\gamma$
will be precisely measured, this multitude of $R_i(\beta,\gamma)$ will provide very good tests of the SM. However, using Step 4 of our strategy it is possible
to predict these ratios already now.

The details of the execution of this strategy can be found in \cite{Buras:2022wpw,Buras:2021nns,Buras:2022qip}. In particular analytic expressions
for $R_i(\beta,\gamma)$ and plots for them can be found in 
\cite{Buras:2021nns}. Additional ratios,  predictions for all ratios considered  and  for 26 individual branching ratios resulting
from our strategy are presented in \cite{Buras:2022qip}.
Here we just list few results obtained in these papers.

Presently the most interesting $\vcb$-independent ratios read
\be\label{R13}
  \boxed{\frac{\mathcal{B}(B_s\to\mu^+\mu^-)}{\Delta M_s}=
    (2.13\pm0.07)\times 10^{-10}\text{ps}\,,}
\ee
   \be\label{R14}
\boxed{\frac{\mathcal{B}(B_d\to\mu^+\mu^-)}{\Delta M_d}= (2.02\pm0.08)\times 10^{-10}\text{ps}\,,}
\ee
  
\be\label{R11}
  \boxed{\frac{\mathcal{B}(\kpn)}{|\varepsilon_K|^{0.82}}=(1.31\pm0.05)\times 10^{-8}{\left(\frac{\sin\gamma}{\sin 64.6^\circ}\right)^{0.015}\left(\frac{\sin 22.2^\circ}{\sin \beta}\right)^{0.71},  }            }
  \ee
  \be\label{R12a}
\boxed{\frac{\mathcal{B}(\klpn)}{|\varepsilon_K|^{1.18}}=(3.87\pm0.06)\times 10^{-8}
    {\left(\frac{\sin\gamma}{\sin 64.6^\circ}\right)^{0.03}\left(\frac{\sin\beta}{\sin 22.2^\circ}\right)^{0.9{8}}.}}
\ee

Using the experimental values of $\Delta M_s$, $\Delta M_d$ and $|\varepsilon_K|$
they imply the most accurate predictions for the four branching ratios in question in  the SM  to date \cite{Buras:2022wpw,Buras:2021nns}. Moreover, they are independent of the value of $\vcb$. We find  
\be\label{SM1}
\overline{\mathcal{B}}(B_s\to\mu^+\mu^-)_\text{SM}= (3.78\pm 0.12)\times 10^{-9}\,, \quad
\mathcal{B}(B_d\to\mu^+\mu^-)_\text{SM}=(1.02\pm 0.12)\times 10^{-9},
\ee
and 
\be\label{SM2}
\mathcal{B}(\kpn)_\text{SM}= (8.60\pm 0.42)\times 10^{-11}\,, \quad
\mathcal{B}(\klpn)_\text{SM}=(2.94\pm 0.15)\times 10^{-11}.
\ee
In particular, the uncertainties in the latter two branching ratios
have been reduced relative to widly quoted 2015 values  \cite{Buras:2015qea} by a factor of $2.4$ and $4.0$, respectively. In this context the reduction
of theoretical uncertainties in $\varepsilon_K$ \cite{Brod:2019rzc} was important.

Comparing them with the experimental data
\be\label{EXPK}
\mathcal{B}(\kpn)_\text{SM}= (10.9\pm 3.8)\times 10^{-11}\,, \quad
\mathcal{B}(\klpn)_\text{SM}\le 2.0\times 10^{-9}, 
\ee
from NA62 \cite{NA62:2022hqi} and KOTO \cite{Ahn:2018mvc}, respectively, it is clear that there is still a large room left for NP contributions. However, in
order to identify it, it is crucial to measure both branching ratios with at least $5\%$ accuracy. I hope that NA62 and KOTO collaborations will be supported in
this important goal by CERN and J-PARC authorities.

Also there  is a large room for NP in the case of $B_d\to\mu^+\mu^-$ decay. But in the case of $B_s\to\mu^+\mu^-$ decay
the HFLAV average of CMS, LHCb and ATLAS data reads
\be\label{SMX}
\overline{\mathcal{B}}(B_s\to\mu^+\mu^-)_\text{EXP}= (3.45\pm 0.29)\times 10^{-9}\,, \qquad
(\text{HFLAV})
\ee
and consequently the room left for NP is much smaller. Therefore  the precise value for this branching ratio obtained
by us could one day, when the data improves, help to uncover some NP contributions.

 Using this strategy we obtained SM predictions for 26 branching
 ratios for rare semileptonic and leptonic $K$ and $B$ decays with the $\mu^+\mu^-$ pair   or the $\nu\bar\nu$ pair in the final state. They are listed
 in the Tables in \cite{Buras:2021nns,Buras:2022qip}.  
 \subsubsection{Rapid Tests}
Having set the SM expressions for $\Delta F=2$ observables to their
      experimental values we are now in the position to determine the
      CKM parameters. However, before doing it, it is mandatory to perform the third step of our strategy, namely the
      rapid test to be sure that the $\Delta F=2$ observables and the resulting CKM parameters are not infected
      by NP. To this end, instead of inserting the formulae in a computer
      program right away it is useful to 
      construct first a $\vcb-\gamma$ plot \cite{Buras:2021nns,Buras:2022wpw} with three bands resulting  separately from $\Delta M_s$, $\Delta M_d$ and
      $|\varepsilon_K|$  and in the latter case imposing the
      constraint from $S_{\psi K_S}$.

\begin{figure}[t!]
  \centering%
  \includegraphics[width=0.70\textwidth]{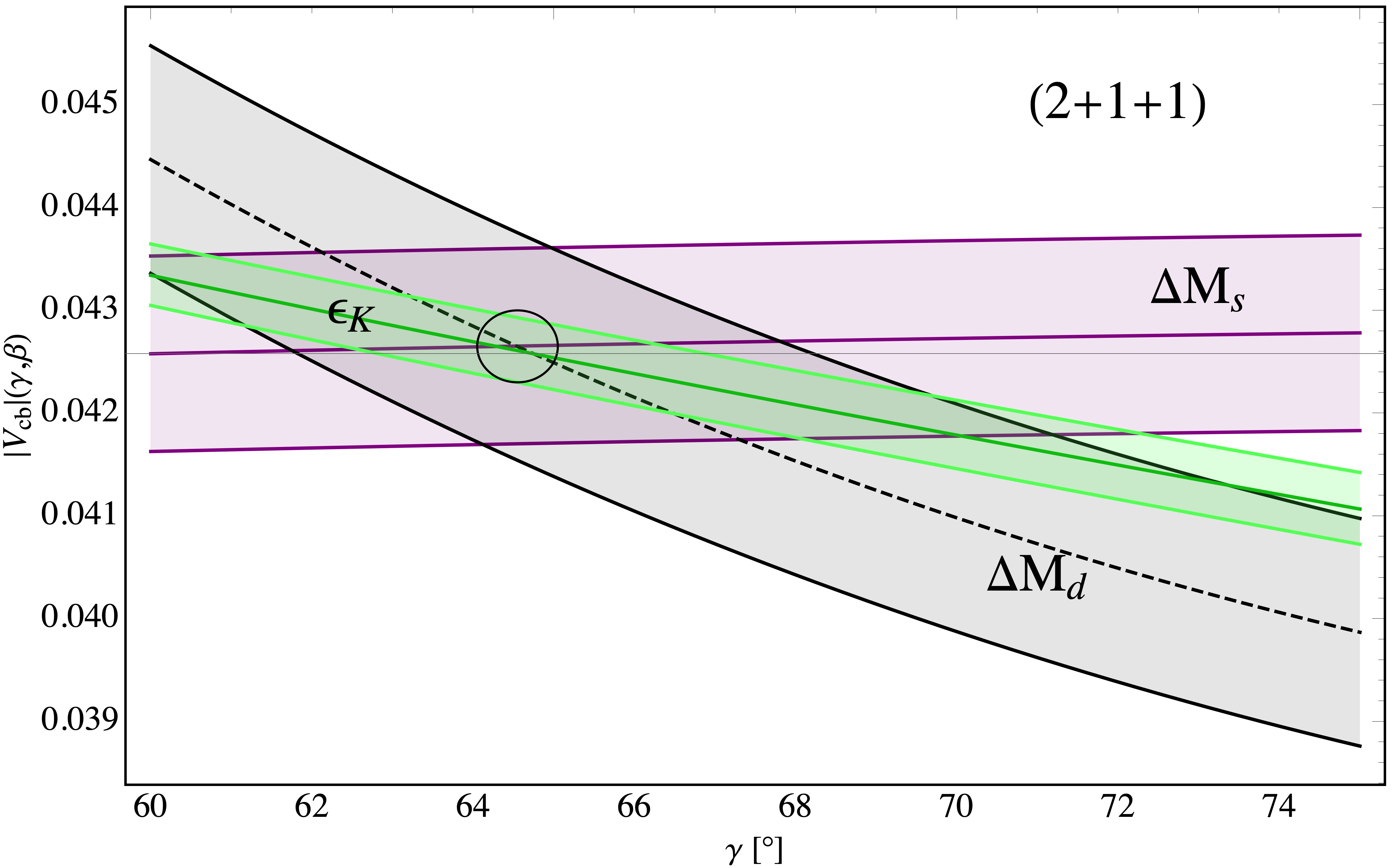}\\
 \includegraphics[width=0.70\textwidth]{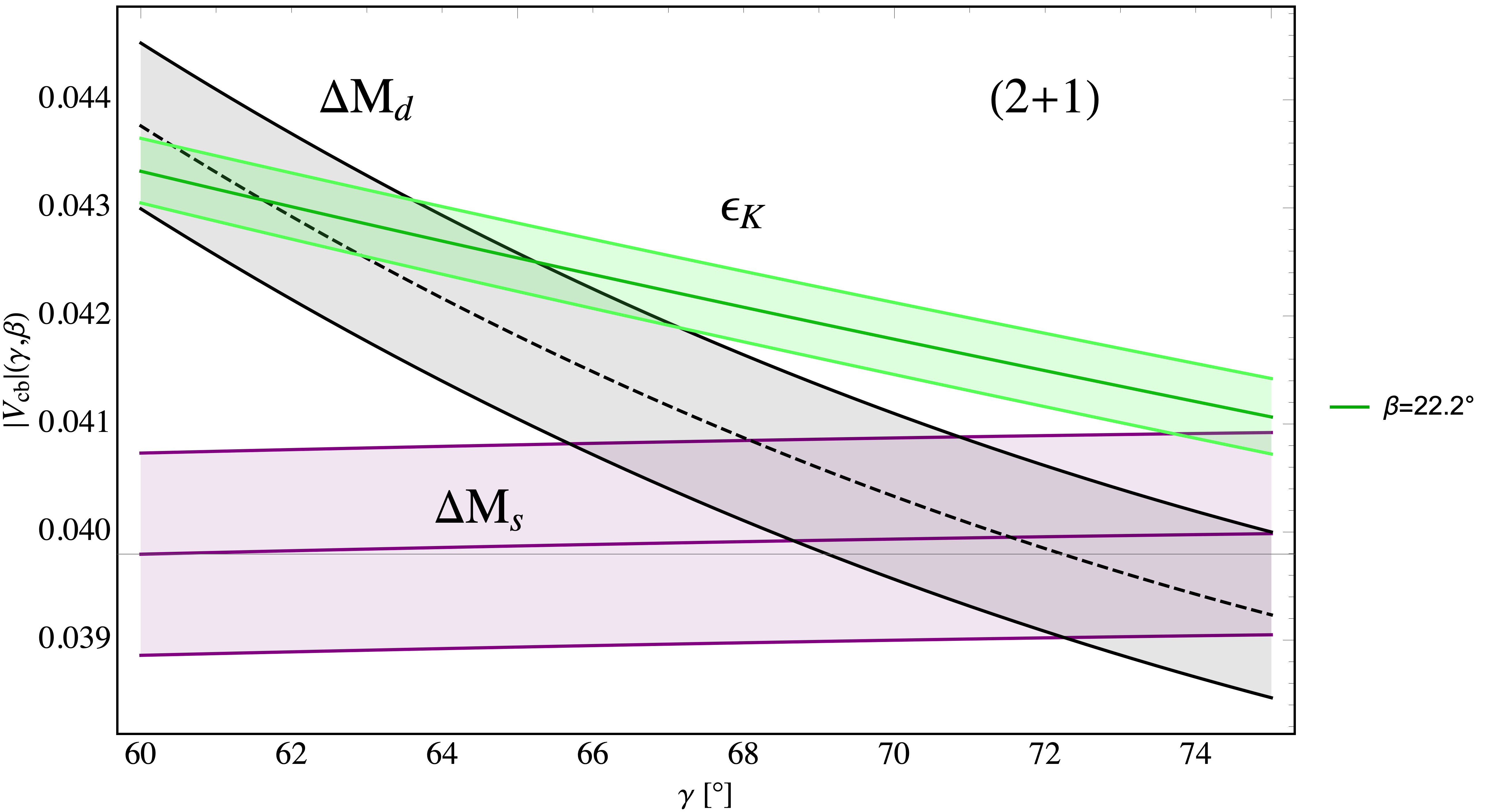}\\
\includegraphics[width=0.70\textwidth]{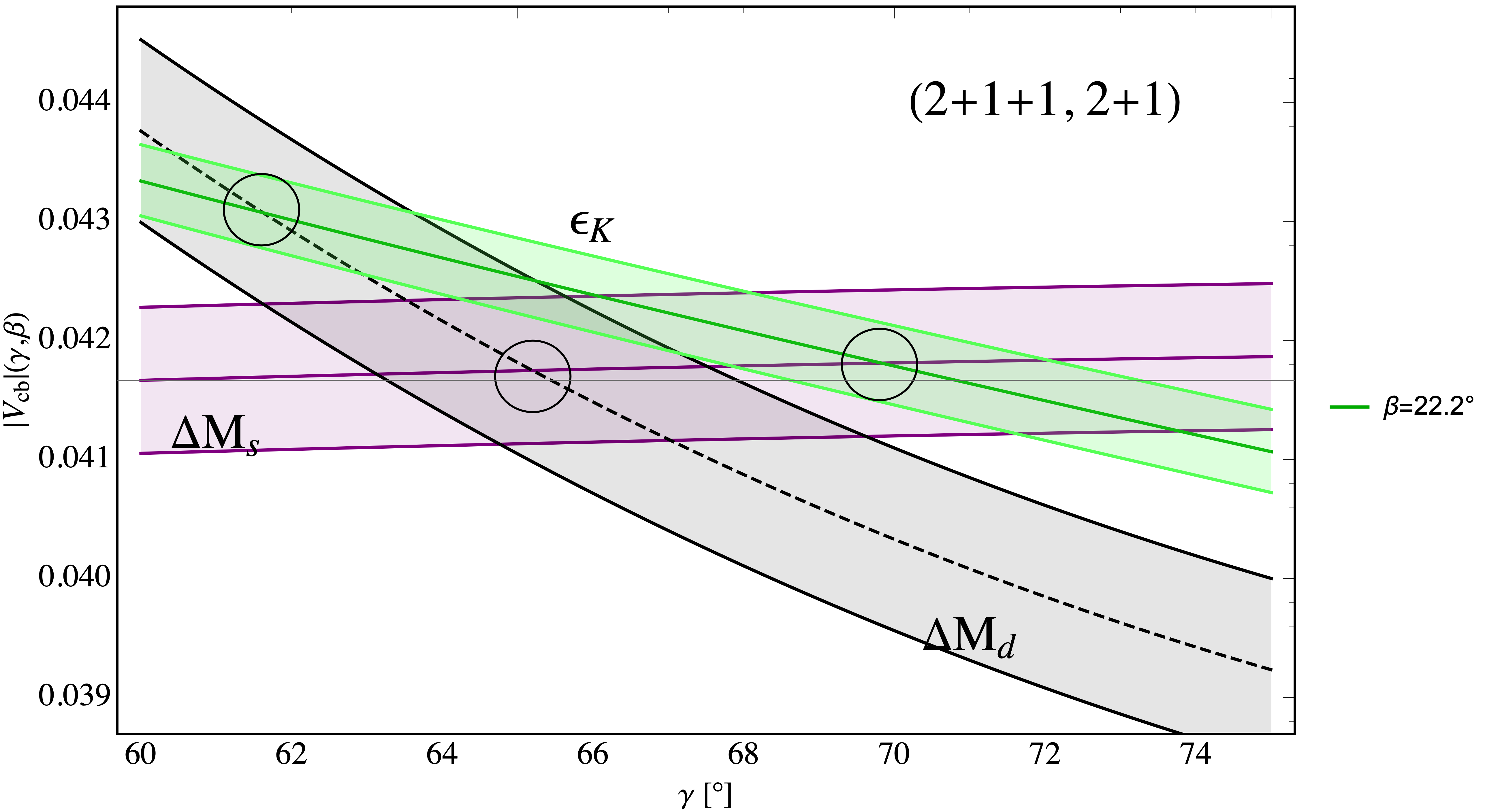}%
\caption{\it {Three rapid tests of NP infection in the $\Delta F=2$ sector taken from \cite{Buras:2022wpw} as explained in the text. The values of $\vcb$ extracted from $\varepsilon_K$, $\Delta M_d$ and  $\Delta M_s$ as functions of $\gamma$. $2+1+1$ flavours (top), $2+1$ flavours (middle), average of $2+1+1$ and $2+1$
    cases (bottom). The green band represents experimental $S_{\psi K_S}$ constraint on $\beta$.}
\label{fig:5}}
\end{figure}

The plots in Fig.~\ref{fig:5}, taken from \cite{Buras:2022wpw}, illustrate
three  {\em rapid tests} of NP infection of the  $\Delta F=2$ sector.
      The test is {\em negative} if these 
      three bands cross each other at a small common area in this plane so
      that unique values of $\vcb$ and $\gamma$ are found. Otherwise it is {\em positive} signalling NP infection. Indeed, 
      as seen in the first $\vcb-\gamma$ plot in  Fig.~\ref{fig:5} that is based on $2+1+1$ LQCD hadronic matrix elements  \cite{Dowdall:2019bea},
      the SM $\vcb-\gamma$ bands resulting from $\varepsilon_K$, $\Delta M_d$ and  $\Delta M_s$ after imposition of the  $S_{\psi K_S}$ constraint, turn out to provide such  unique values of $\vcb$ and $\gamma$. No sign of NP infection in this case. On the other hand, as seen in the remaining
      two plots in   Fig.~\ref{fig:5},  this is not the case if $2+1$
      or the average of $2+1+1$ and $2+1$ hadronic matrix elements LQCD are used. In these two cases the test turns out to be {\em positive}.

      The superiority of the  $\vcb-\gamma$ plots in general 
      with respect to $\vcb$ and  $\gamma$ over UT plots
      has been emphasized in  \cite{Buras:2022nfn}. Indeed,
\begin{itemize}
\item
  They exhibit $\vcb$ and its correlation with $\gamma$ determined through a given observable in the SM, allowing thereby monitoring the progress 
  on both parameters expected in the coming years. Violation of this correlation in experiment will clearly indicate NP at work.
\item
  They utilize the strong sensitivity of rare $K$ decay  processes to $\vcb$ thereby providing
  precise determination of $\vcb$
  even with modest experimental precision on their branching ratios.
\item
  They exhibit, as shown below, the action of $\Delta M_s$ and of $B_s$ decays,
like $B_s\to\mu^+\mu^-$
which is not possible in the common UT-plot.
\item
  Once the accuracy of $\gamma$ measurements  will approach $1^\circ$ it will be easier to monitor this progress on a $\vcb-\gamma$ plot.
\end{itemize}

\begin{figure}[t!]
  \centering%
  \includegraphics[width=0.70\textwidth]{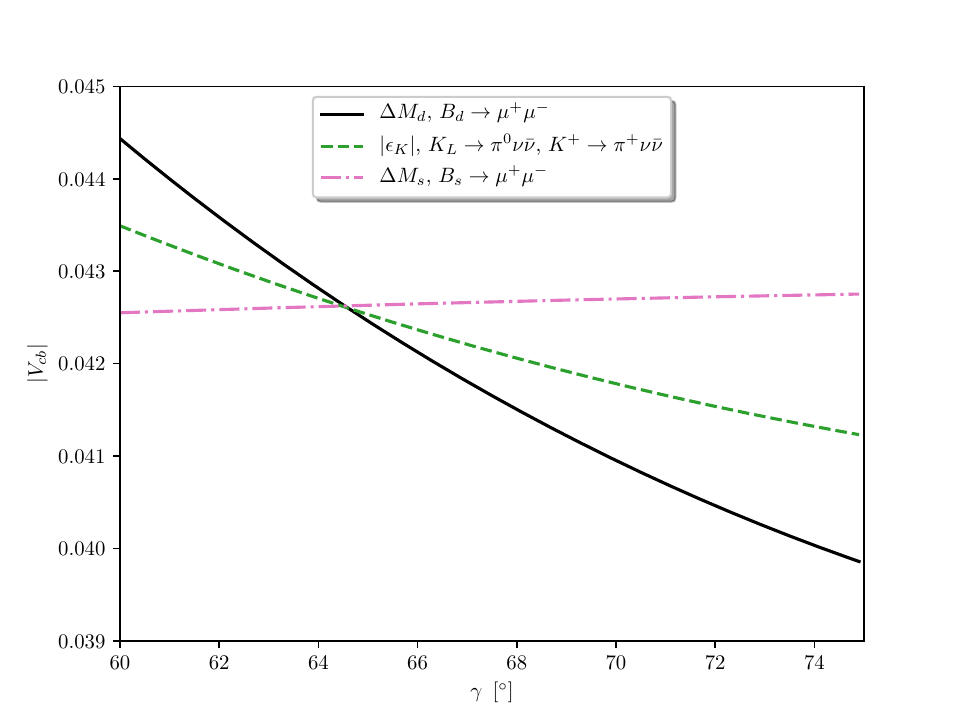}
  \caption{\it {Schematic illustration of the action of the seven observables in the $\vcb-\gamma$ plane in the context of the SM. We set $\beta=22.2^\circ$ and all uncertainties
      to zero.}
\label{fig:X}}
\end{figure}

In order to illustrate this
we show in Fig.~\ref{fig:X} the results for a number of observables calculated in the SM setting all uncertainties
for transparency reasons to zero. We make the following observations.
\begin{itemize}
\item
  For fixed $\beta=22.2^\circ$, $\varepsilon_K$, $\kpn$ and $\klpn$ are represented to an excellent approximation by the same line which is already a very good test of the SM. This is simply because  the $\gamma$ dependence in the three observables is practically the same, the fact pointed out first in  \cite{Buchalla:1994tr} and
  strongly emphasized in \cite{Buras:2021nns}.
The dependence
on $\beta$ is different and this allows to determine within the SM  the angle $\beta$ from any pair of these observables
independently of the value of $\gamma$.
For the pair of the rare $K$ branching ratios this was pointed out in
  \cite{Buchalla:1994tr}. For the other two pairs in \cite{Buras:2021nns}.
 \item
  $\Delta M_d$ and $B_d\to\mu^+\mu^-$ are represented by a single line
  and a different line represents $\Delta M_s$ and  $B_s\to\mu^+\mu^-$.
  This is precisely the illustration of the SM relations  (\ref{R13}) and
(\ref{R14})   pointed
  out long time ago in \cite{Buras:2003td}.
\end{itemize}
\begin{figure}[t!]
  \centering%
  \includegraphics[width=0.70\textwidth]{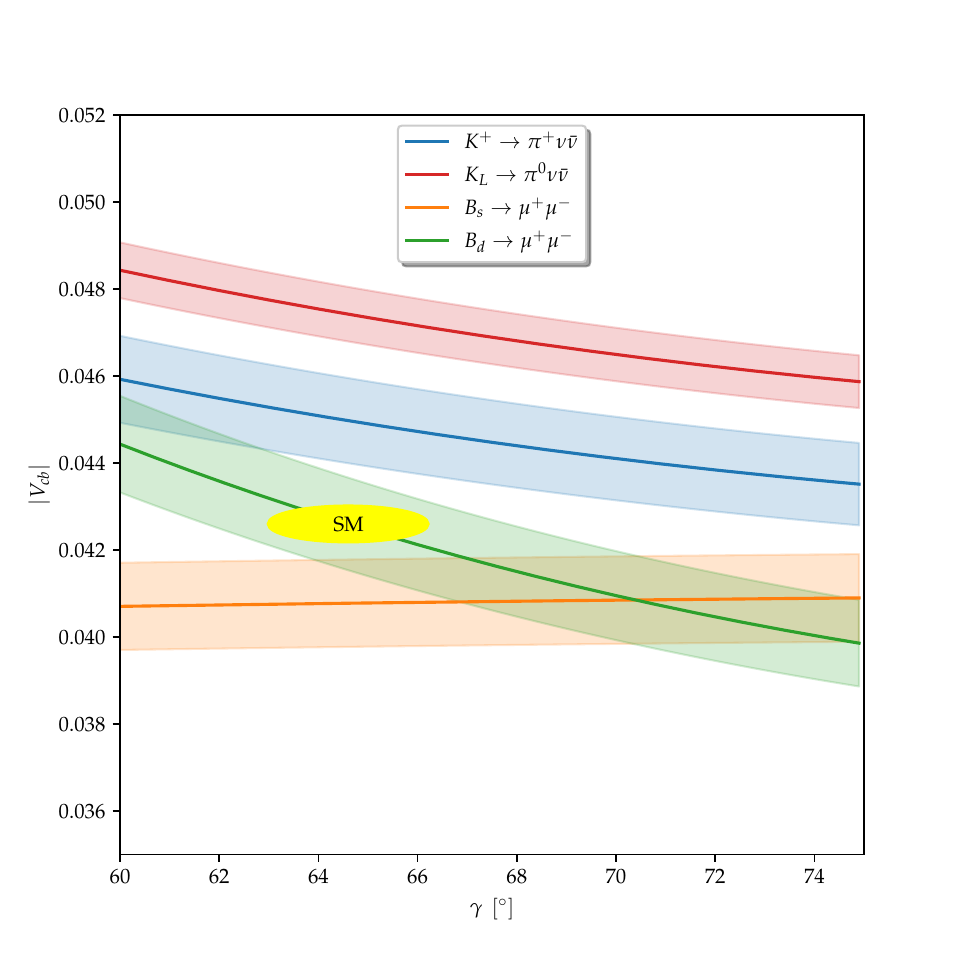}
  \caption{\it {The impact of hypothetical future measurements of 
      the branching ratios for $\kpn$, $\klpn$,  $B_d\to\mu^+\mu^-$ and $B_s\to\mu^+\mu^-$ 
       on
      the $\vcb-\gamma$ plane. All uncertainties  are included. The yellow disc
    represents the SM as obtained in {(\ref{CKMoutput})}.}
\label{fig:Y}}
\end{figure}

While SM describes  $\varepsilon_K$,
$\Delta M_d$, $\Delta M_s$ simultaneously very well, this not need to be the case for the
 rare decays in question. This is illustrated  in Fig.~\ref{fig:Y}. To obtain these results we have
set the branching ratio for $B_s\to\mu^+\mu^-$ to
the experimental world average from LHCb, CMS and ATLAS \cite{LHCb:2021awg,CMS:2020rox,ATLAS:2020acx}  but decreased its error from $8.4\%$ down to $5\%$. For the remaining branching ratios we have chosen values resulting from hypothetical future measurements that differ from the SM predictions.
We kept the errors at $5\%$ as in the case of $B_s\to\mu^+\mu^-$ to exhibit the superiority  of rare $K$ decays  over rare $B$ decays as far as the determination of $\vcb$ is concerned.  
While the experimental errors are futuristic, we expect that the theoretical
errors will go down with time so that the bands in Fig.~\ref{fig:Y} could
apply one day with less accurate measurements.

This plot confirms all the statements made above. The superiority of
$\klpn$ over the remaining decays is clearly seen. The blue band will be narrowed once the long distance charm contributions to $\kpn$ will be known with higher precision 
from lattice QCD calculations \cite{Christ:2019dxu} than they are known now
\cite{Isidori:2005xm}.

\subsubsection{CKM Parameters}
As the rapid test for the $\Delta F=2$ observables turned out to be negative we can now determine the CKM parameters using these observables without 
NP infection. We find \cite{Buras:2022wpw}
\be\label{CKMoutput}
\boxed{\vcb=42.6(4)\times 10^{-3}, \quad 
\gamma=64.6(16)^\circ, \quad \beta=22.2(7)^\circ, \quad \vub=3.72(11)\times 10^{-3}\,}
\ee
and consequently
\be\label{CKMoutput2}
\boxed{\vts=41.9(4)\times 10^{-3}, \qquad \vtd=8.66(14)\times 10^{-3}\,,\qquad
{\IM}\lambda_t=1.43(5)\times 10^{-4}\,,}
\ee
\be\label{CKMoutput3}
\boxed{\bar\varrho=0.164(12),\qquad \bar\eta=0.341(11)\,,}
\ee
where $\lambda_t=V_{ts}^*V_{td}$.

The hierarchy
\be
|V_{td}V_{ts}^*|\approx 3.6\times 10^{-4}, \qquad |V_{td}V_{tb}^*|\approx 8.7\times 10^{-3},\qquad |V_{ts}V_{tb}^*|\approx 41.9\times 10^{-3}
\ee
implies a hierarchy in FCNC processes in $K$, $B_d$ and $B_s$ meson systems.
The strong suppression of such processes in the $K$ system in the SM allows
in principle larger NP effects than in the $B_d$ and $B_s$ meson decays.

\subsubsection{More Results}
Having the values of the CKM parameters, all ratios $R_i(\beta,\gamma)$ considered by us can be predicted in the SM. We quote here only  \cite{Buras:2022qip}
 \be\label{R1}
\boxed{\frac{\mathcal{B}(\kpn)}{\left[{\overline{\mathcal{B}}}(B_s\to\mu^+\mu^-)\right]^{1.4}}= 53.69\pm2.75\,,}
\ee
\be\label{R5}
\boxed{\frac{\mathcal{B}(\kpn)}{\left[\mathcal{B}(B^+\to K^+\nu\bar\nu)\right]^{1.4}}=(1.94\pm0.13)\times 10^{-3}\,.}
\ee

Of particular interest are also SM predictions for the 
$B^+\to K^+\mu^+\mu^-$  and $B_s\to \phi\mu^+\mu^-$ branching ratios. In
the low $q^2$ bin they imply the pulls ($-4.4\sigma$) and ($-4.8\sigma$), respectively \cite{Buras:2022qip}.  Both branching ratios are measured to be
suppressed relative to the SM predictions. To my knowledge these are the largest
anomalies in single branching ratios found in the literature to date. The
reason are not only more precise values of CKM parameters used by us than in
other papers, but also the most recent form factors from HPQCD lattice collaboration \cite{Parrott:2022rgu,Parrott:2022zte}.

These findings  together with the strong suppression of NP to $\Delta F=2$
observables 
put very strong constraints on NP models attempting to explain these
$B$ physics anomalies and  other anomalies like the one in the ratio $\epe$.
Indeed, despite some controverses, it is likely that the SM prediction for $\epe$ has to be enhanced by NP
  to agree with data \cite{Buras:2022cyc}. Also the  most recent SM results for $\Delta M_K$ from  the RBC-UKQCD lattice
  collaboration \cite{Bai:2018mdv,Wang:2022lfq}  are significantly larger than its very precise experimental value although
  due to large uncertainties this deviation is only around $2.0\,\sigma$.
 
The question then arises which NP could explain these anomalies without
destroying good agreement of the SM with experimental  data on the
$\Delta F=2$ observables. This is the subject of the 3rd
movement of this KM symphony.

\boldmath
\subsection{$Z^\prime$ at Work}\label{Zprime}
\unboldmath
\boldmath
\subsubsection{Kaon Physics without New Physics in $\varepsilon_K$}\label{MBABK}
\unboldmath
Concentrating on the $K$ system,  one could at first sight start worrying
that the absence of NP in a CP-violating observable like $\varepsilon_K$
would exclude all NP effects in rare decays governed by CP violation
 such as $\klpn$, $K_S\to\mu^+\mu^-$, $K_L\to\pi^0\ell^+\ell^-$ and also in the ratio $\epe$. Fortunately, these worries are premature because  $\varepsilon_K$ is governed by CP violation in mixing while the remaining observables are either fully dominated by CP violation in decay (direct CP violation) or significantly affected by it. Indeed,
as pointed out already in 2009, in an important paper by Monika Blanke
\cite{Blanke:2009pq}, the absence of NP in $\varepsilon_K$ does not preclude
the absence of NP in these observables. This follows from the simple fact that
\be\label{EPSILONNEW}
(\varepsilon_K)_\text{BSM}\propto\left[\RE(g_{sd})\IM(g_{sd})\right],
\ee
where $g_{sd}$ is a  complex coupling present in a given NP model. Setting
$\RE(g_{sd})=0$, that is making this coupling {\em imaginary}, eliminates NP {contributions to} $\varepsilon_K$, while still allowing for sizable CP-violating effects in
rare decays and $\epe$. {This choice automatically eliminates the second solution considered in \cite{Blanke:2009pq} ($\IM(g_{sd})=0$), which is clearly less interesting.}

\begin{figure}[t]
\centering%
\includegraphics[width=0.6\textwidth]{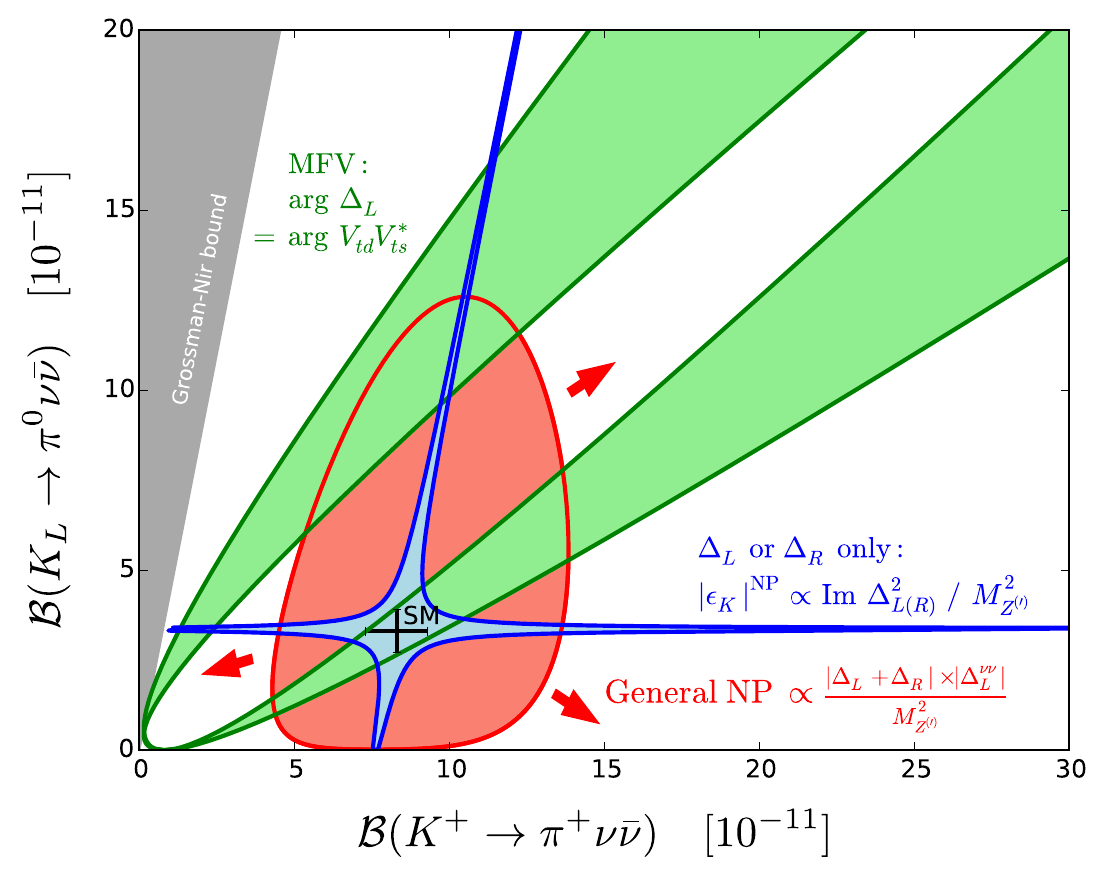}%
\caption{\it Illustrations of common correlations in the $\mathcal{B}(\kpn)$ versus $\mathcal{B}(\klpn)$ plane. The expanding red region illustrates the lack of correlation for models with general LH and RH NP couplings. The green region shows the correlation present in models obeying Constrained Minimal Flavour Violation (CMFV). The blue region shows the correlation induced by the constraint from $\varepsilon_K$ if only LH or RH couplings  are present. From \cite{Buras:2015yca}.  \label{fig:illustrateEpsK}}
\end{figure}

These two solutions are  exhibited in Fig.~\ref{fig:illustrateEpsK} through
two blue branches on which the correlation between two branching ratios takes
place when only LH or RH couplings are present and NP contributions to 
$\varepsilon_K$ are strongly suppressed. More details on this figure can be
found in \cite{Buras:2015yca} where various simplified models have been considered.

\begin{figure}[t]
\begin{center}
 \includegraphics[width=1.\textwidth]{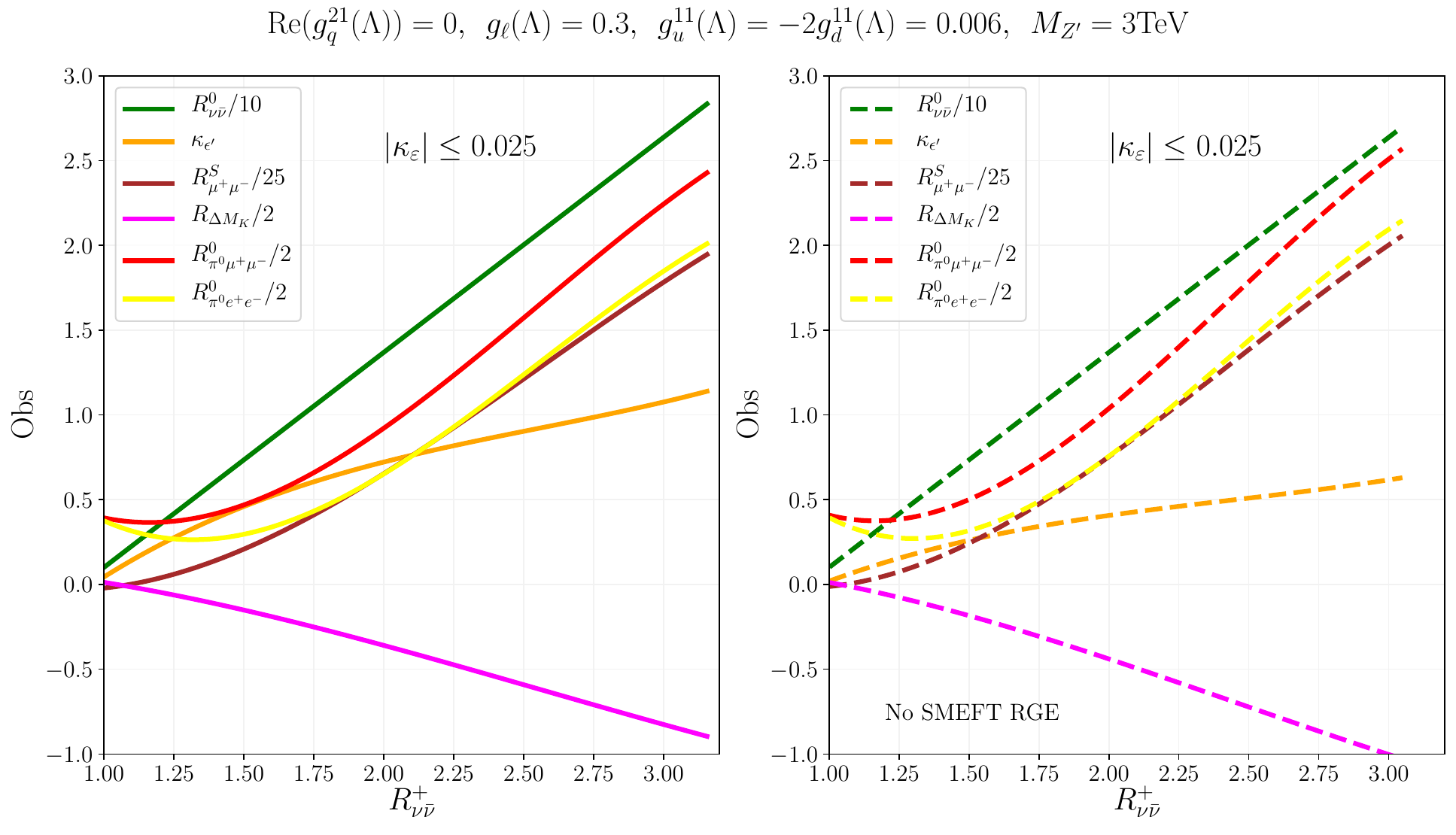}
\captionsetup{width=0.9\textwidth}
\caption{The correlations between the $\kpn$ ratio
$R_{\nu \bar \nu}^+$ and various other ratios for other  Kaon decays. All ratios are equal unity in the SM.  From 
  \cite{Aebischer:2023mbz}.}
\label{Fig:2}
\end{center}
\end{figure}

A detailed analysis of $K$ physics without NP  in $\varepsilon_K$   in a simple $Z^\prime$ model has been presented
recently in \cite{Aebischer:2023mbz}. It has been demonstrated that indeed
 significant NP contributions to $\kpn$, $\klpn$, $K_S\to\mu^+\mu^-$, $K_L\to\pi^0\ell^+\ell^-$, $\epe$ and $\Delta M_K$
 can be present despite no NP contributions to $\varepsilon_K$. This {scenario} implies very stringent correlations between {the} Kaon observables considered by us. In particular, the identification of NP in any of these
  observables implies automatically NP contributions to the remaining ones
  under the assumption of non-vanishing flavour conserving $Z^\prime$ couplings
  to $q\bar q$, $\nu\bar\nu$, and $\mu^+\mu^-$. A characteristic
  feature of this scenario is a strict correlation between   $\kpn$ and $\klpn$ branching ratios on a branch parallel to the Grossman-Nir bound \cite{Grossman:1997sk} pointed
  out in \cite{Blanke:2009pq} and therefore usually called Monika Blanke (MB) branch.   Moreover, $\Delta M_K$ is automatically suppressed as it seems to be required by the results of the RBC-UKQCD lattice QCD collaboration \cite{Bai:2018mdv,Wang:2022lfq}. Furthermore, there is no NP contribution to $K_L\to\mu^+\mu^-$ which otherwise would bound NP effects in $\kpn$. Of particular interest are the correlations of $\kpn$ and $\klpn$ branching ratios and   of $\Delta M_K$  with the ratio $\epe$. These correlations are summarized in Fig.~\ref{Fig:2},
  where we show the ratios of total branching ratios to the SM ones as the function of such ratio for $\kpn$. 
  All ratios shown there are equal unity in the SM. $R^+_{\nu\bar\nu}$ should be measured within
  $10\%$ accuracy by NA62 collaboration in the coming years and hopefully with
  the accuracy of $5\%$ still in this decade. Similar comments apply to $R^0_{\nu\bar\nu}$ measured by KOTO. Note that for $R^+_{\nu\bar\nu}=1.75$ we predict 
  $R^0_{\nu\bar\nu}=10$ but one should realize that in a different NP scenario these correlations could be different.

The parameter $\kepe$ is defined as follows
\cite{Buras:2015jaq}
\be\label{GENERAL}
\frac{\varepsilon'}{\varepsilon}=\left(\frac{\varepsilon'}{\varepsilon}\right)^{\rm SM}
+\left(\frac{\varepsilon'}{\varepsilon}\right)^{\rm BSM}\,, \qquad
\left(\frac{\varepsilon'}{\varepsilon}\right)^{\rm BSM}= \kepe\cdot 10^{-3}\,, \qquad   0.0 \le \kepe \le 1.2\,.
\ee

In the case of $\varepsilon_K$ we allow only for very small NP contributions that could be generated by RG effects despite setting the real part of
$g_{sd}(Z^\prime)$ to zero at the NP scale that we take to be equal
to $ M_{Z^\prime}$. Explicitly
\be
(\varepsilon)^{\rm BSM}=\kappa_\varepsilon\cdot  10^{-3}\,, \qquad -0.025\le |\kappa_\varepsilon|\le 0.025\,,
\ee
which amounts to $1\%$ of the experimental value. 

It should be noted that some ratios are enhanced so much that in order to show
the results in one plot they had to be divided by a suitable factor. This is in particular the case of $\klpn$ and $K_S\to\mu^+\mu^-$.

\boldmath
\subsubsection{B-Meson Physics without New Physics in $B_{s,d}^0-\bar B_{s,d}^0$
Mixing}
\unboldmath
The strategy just outlined 
does not eliminate tree-level NP contributions to $\Delta M_s$ and $\Delta M_d$  because they  are governed
by the absolute values of the mixing amplitudes ($ij$ are quark flavours)
\be
M^{ij}_{12}= (M^{ij}_{12})^{\rm SM}+M^{ij}_{12}(Z^\prime),
\ee
with $ij=bs,~bd$
and  not by their imaginary parts as is the case of $\varepsilon_K$. 
Therefore
in order  to remove NP contributions to observables in $B^0_{s,d}-\bar B^0_{s,d}$
systems at tree-level, we have to remove $M^{ij}_{12}(Z^\prime)$ completely at this level, while keeping the $Z^\prime bs$ and $Z^\prime bd$ couplings non-zero which
is required for the explanation of the observed anomalies in $b\to s\mu^+\mu^-$
transitions.

This is not possible with a single $Z^\prime$ gauge boson with only flavour-violating left-handed couplings. At present it appears that there are three possible strategies to solve this problem.
\begin{itemize}
\item
  Allow NP to enter the  $B^0_{s,d}-\bar B^0_{s,d}$ systems at the $5\%$ level
  and to lower a bit the value of $\vcb$ relative to the one in
  (\ref{CKMoutput}). This is necessary in the 331 models as demonstrated recently in \cite{Buras:2023ldz}. The remarkable feature of some of these models
  is the prediction of $\Delta C_9=-b \Delta C_{10}$ with $2\le b\le 5$ so that after most recent LHCb results for $b\to s \ell^+\ell^-$ decays \cite{LHCb:2022qnv,LHCb:2022zom} and CMS result for
  $B_s\to\mu^+\mu^-$, the present $B$ physics anomalies can be  explained  but  NP effects in rare $K$ decays and $\epe$ turn out to be small.
\item
  Next, a  recent suggestion made in  \cite{Buras:2023xby}.
  Including a second $Z^\prime$ in the model allows to eliminate tree-level
  contributions to all $\Delta F=2$  observables while allowing for
  significant tree-level contributions to $K$, $D$ and $B$ decays, including
  CP-violating ones. This allows to explain the existing anomalies in $b\to s\mu^+\mu^-$ transitions and the anticipated anomaly in the ratio $\epe$ much easier than in $Z^\prime$-Single scenarios because one does not have to worry about
  constraints from the $B^0_{s,d}-\bar B^0_{s,d}$ mixings. This strategy can also be used not only for $B^0_{s,d}-\bar B^0_{s,d}$ mixings but also for $K^0-\bar K^0$ and $D^0-\bar D^0$ systems if this turns out to be required.
  However, while
  just looking at Feynman diagrams these cancellations of NP contributions
  to $\Delta F=2$ processes can be easily arranged as described below, it
  is presently unclear whether such a construction can follow from
  a UV completion and possibly more $Z^\prime$ gauge bosons than just two
  are required.
\item
  If the single $Z^\prime$ has both left-handed and right-handed flavour-violating quark couplings an interplay of the left-left, right-right and left-right
  $Z^\prime$  contributions can provide necessary suppression of NP to $\Delta F=2$ observables \cite{Buras:2014sba,Buras:2014zga,Crivellin:2015era}.
This requires some 
 fine tuning between these three contributions.
    Fortunately, in the case of $B$ physics this tuning is much smaller
    than in the case of $K$ physics, where this tuning is not required to remove
    NP from $\varepsilon_K$ 
  as demonstrated above \cite{Aebischer:2023mbz}.
\end{itemize}

Let us briefly discuss the second and the third strategy.

\paragraph{$Z^\prime$-Tandem}

  The proposed
  $Z^\prime$-Tandem mechanism for the elimination of NP contributions to
  $B^0_{s,d}-\bar B^0_{s,d}$ mixing \cite{Buras:2023xby}
  bears some similarities to the GIM mechanism
  for the suppression of the FCNCs in the SM  with the role of the
  charm quark played here by the second $Z^\prime$. However, it differs from the latter profoundly in that only NP contributions to quark mixing are eliminated at tree-level, while GIM mechanism removes them from all FCNC processes at tree-level.

 Denoting then the quark couplings of these two gauge bosons by
  \be
\Delta_L^{ij}(Z_1^\prime)=|\Delta_L^{ij}(Z_1^\prime)|e^{i\phi^{ij}_1}, \qquad
\Delta_L^{ij}(Z_2^\prime)=|\Delta_L^{ij}(Z_2^\prime)|e^{i\phi^{ij}_2}\,,
\ee
where $(i,j)$ are quark flavour indices, either for down-quarks or up-quarks,
 the two conditions for the removal of the $Z^\prime_{1,2}$ contributions to
$M^{ij}_{12}$ at tree-level read as follows
  \be\label{c1}
 \boxed{ \frac{|\Delta_L^{ij}(Z_1^\prime)|}{M_1}=\frac{|\Delta_L^{ij}(Z_2^\prime)|}{M_2}\,,
  \qquad
  \phi^{ij}_2=\phi^{ij}_1+90^\circ\,,}
  \ee
 with $M_{1,2}$ being the masses of $Z^\prime_{1,2}$.

The following comments should be  made.
\begin{itemize}
\item
  The cancellation in question implies the presence of new CP-violating phases
  which will be visible in $B$, $K$ and $D$ decays. The phases $\phi^{ij}_1$ and  $\phi^{ij}_2$ cannot all vanish simultaneously.
\item In case of some signs of NP contributions to $\Delta F=2$, the conditions in (\ref{c1}) could be relaxed. This could turn out to be the case of $\Delta M_K$ analyzed in a $Z^\prime$-Single scenario in \cite{Aebischer:2023mbz}.
\end{itemize}

  The implied flavour patterns in  $K$ and $B$ decay
  observables in this NP   scenario have been only briefly discussed in \cite{Buras:2023xby}. Here we just mention
  one difference relative to the single $Z^\prime$ scenario.
  The   correlation of $\kpn$ and $\klpn$ branching ratios is predicted to take place {\em outside} the MB branch. If this turned out to be the case one day
  and no sign of NP in $\Delta F=2$ processes would be observed, this
  would be a hint for $Z^\prime$-Tandem mechanism at work, in particular
  if the lighter $Z^\prime$ gauge boson had been discovered already.

  Let us summarize the main structure of $Z^\prime$-Tandem framework proposed in
\cite{Buras:2023xby} 
\begin{itemize}
\item
  The tandem collaborates to remove NP from quark mixing because
  NP is not required to fit data. In this manner CKM matrix can be
  determined without NP infection. 
\item
  The NP parameters are then determined exclusively from deviations of
  experimental results for $B$, $K$ and $D$
  decays from SM predictions without any worry about $\Delta F=2$ constraints. 
  Moreover, only tree-level NP contributions have to be considered because
  of a very strong suppression of one-loop contributions in this scenario.
\end{itemize}
Yet, although I mentioned this idea here and I was excited about this solution while writing \cite{Buras:2023xby}, my excitement decreased by now because it is difficult,
if not impossible, to construct a UV completion which would imply such a $Z^\prime$-Tandem and satisfy the conditions in (\ref{c1}). Possibly increasing the
number of $Z^\prime$ gauge bosons  would be helpful but I leave this for the future and will now describe the simplest solution.

\paragraph{Right-handed Couplings at Work}

Including both left-handed and right-handed couplings of $Z^\prime$ to quarks one finds the following
shifts in the one-loop functions governing  $B_d^0-\bar B_d^0$ and $B_s^0-\bar B_s^0$  mixings \cite{Buras:2014zga}
\be\label{Sd}
\Delta S(B_d)=2.38
\left[\frac{\Delta_L^{bd}(Z^\prime)}{V_{td}V^*_{tb}}\right]^2\left(\frac{3\teV}{M_{Z^\prime}}\right)^2z^{bd}\,,
\ee
\be\label{Ss}
\Delta S(B_s)=2.38
\left[\frac{\Delta_L^{bs}(Z^\prime)}{V_{ts}V^*_{tb}}\right]^2\left(\frac{3\teV}{M_{Z^\prime}}\right)^2z^{bs}\,,
\ee
where
\be\label{ZprimeMAIN}
z^{bq}=\left[1+\left(\frac{\Delta_R^{bq}(Z^\prime)}{\Delta_L^{bq}(Z^\prime)}\right)^2+2\kappa_{bq}\frac{\Delta_R^{bq}(Z^\prime)}{\Delta_L^{bq}(Z^\prime)}\right],\qquad \kappa_{bq}=\frac{\langle \hat Q_1^\text{LR}(M_{Z^\prime})\rangle^{bq}}{\langle \hat Q_1^\text{VLL}(M_{Z^\prime})\rangle^{bq}}.
\ee
Here $\kappa_{bq}\approx -5$ is the ratio of left-left and left-right handronic
matrix elements. It follows that for
\be\label{LHRH}
\Delta_R^{bq}(Z^\prime)\approx 0.1\,\Delta_L^{bq}(Z^\prime)
\ee
$Z^\prime$ contributions to $B_d^0-\bar B_d^0$ and $B_s^0-\bar B_s^0$  mixings
will be strongly suppressed. Simultaneously, the presence of the right-handed
couplings will have some impact on rare $B$ decays. In the case of
$K^0-\bar K^0$ mixing the corresponding condition reads
\be
\Delta_R^{sd}(Z^\prime)\approx 0.004\,\Delta_L^{bq}(Z^\prime),
\ee
implying large fine tuning required to remove NP contribution but then
also negligible impact of right-handed currents on rare $K$ decays.
Here the solution proposed in  \cite{Aebischer:2023mbz} should be preferred
unless eventually the SM $\Delta M_K$ will agree with experiment.

The implications of this  suppression of NP to $B_d^0-\bar B_d^0$ and $B_s^0-\bar B_s^0$  mixings
can be tested in $b\to s\nu\bar\nu$
transitions. Using our strategies together with $B\to K$ form factor from HPQCD \cite{Parrott:2022zte,Parrott:2022rgu,Parrott:2022smq} we find first \cite{Buras:2022qip}
\be\label{BVNEW}
{\mathcal{B}}(B^+\to K^+\nu\bar\nu)_{\rm SM} =(5.59\pm 0.31)\times 10^{-6},\qquad
{\mathcal{B}}(B^0\to K^{0*}\nu\bar\nu)_{\rm SM} = (10.13\pm 0.92)\times 10^{-6}.
\ee
The main uncertainty comes then from hadronic form factors that should
be known with improved accuracy in coming years. However, we should remark that the result for $B^+\to K^+\nu\bar\nu$
includes $10\%$ upward shift from a tree-level contribution pointed out in \cite{Kamenik:2009kc}. Otherwise it would be $(4.98\pm 0.31)\times 10^{-6}$. In fact
the latter result should be comapred with the very recent result from Belle II:
$(24\pm7)\times 10^{-6}$.

On the other hand the best current experimental bounds~\cite{Olive:2016xmw,Grygier:2017tzo} are set by the Belle collaboration 
\begin{align}
{\mathcal{B}}(B^+\to K^+\nu\bar\nu) &=(24\pm7)\times 10^{-6}~,\\
{\mathcal{B}}(B^0\to K^0\nu\bar\nu) &\leq 2.6\times 10^{-5}\quad \text{@ 90\% CL}~,\\
{\mathcal{B}}(B^+\to  K^{+*} \nu\bar\nu) &\leq 4.0\times 10^{-5}\quad \text{@ 90\% CL}~,\\
{\mathcal{B}}(B^0\to K^{0*}\nu\bar\nu) &\leq 1.8\times 10^{-5}\quad \text{@ 90\% CL}\,.
\end{align}

Defining
\be\label{RKRK*}
\mathcal{R}_{K\nu\nu}=\frac{\mathcal{B}(B \to K \nu \bar\nu)}{\mathcal{B}_{\rm SM}(B \to K \nu\bar\nu)} \, , \quad \mathcal{R}_{K^*\nu\nu}=\frac{\mathcal{B}(B \to K^* \nu\bar\nu)}{\mathcal{B}_{\rm SM}(B \to K^* \nu\bar\nu)} \,,
\ee
we have \cite{Grossman:1995gt,Melikhov:1998ug,Altmannshofer:2009ma,Buras:2014fpa}
\begin{align}
 \mathcal{R}_{K\nu\nu}  & = (1 - 2\,\eta)\epsilon^2
 \,, &
 \mathcal{R}_{K^*\nu\nu}
  & =
  (1 +  \kappa_\eta \eta)\epsilon^2
  \,, &
 \mathcal{R}^\nu_{F_L} \equiv \frac{F_L}{F_L^\text{SM}} 
 & =  
  \frac{1+2\eta}{1+\kappa_\eta\eta}
 \,,
\label{eq:epseta-R1}
\end{align}
where
\begin{equation}  \label{eq:epsetadef1}
 {\epsilon = \frac{\sqrt{ |C^\nu_L|^2 + |C^\nu_R|^2}}{|(C^\nu_L)^\text{SM}|}~, \qquad
 \eta = \frac{-\text{Re}\left(C^\nu_L C_R^{\nu *}\right)}{|C^\nu_L|^2 + |C^\nu_R|^2}~,}
\end{equation}
\noindent
such that $\epsilon>0$ and $\eta$ lies in the range $[-\frac{1}{2},\frac{1}{2}]$. $\epsilon=1$ in the SM and $\eta\neq 0$ signals the presence of right-handed currents.  Presently  $\kappa_\eta=1.33\pm0.05$. $F_L$ is the $K^*$ longitudinal polarization fraction. For recent extensive analyses of dineutrino modes see
\cite{Bause:2021cna,He:2021yoz,Bause:2022rrs,Becirevic:2023aov,Bause:2023mfe,Allwicher:2023syp,Dreiner:2023cms}.

The correlations between the ratios $\mathcal{R}_{K\nu\nu}$ and $\mathcal{R}_{K^*\nu\nu}$ with the corresponding ratios with $\mu\bar\mu$ for  various $Z^\prime$ couplings  can be found in Figure~5 of  \cite{Buras:2014fpa}. Due to the dominance  of left-handed $Z^\prime$ couplings  over right-handed ones, as given in (\ref{LHRH}) 
in the present case, one finds  anti-correlation between $\nu\bar\nu$ and $\mu\mu$ channels. The observed suppression of  $\mathcal{R}_{K\mu\mu}$ and $\mathcal{R}_{K^*\mu\mu}$ below unity implies then the enhancements of  $\mathcal{R}_{K\nu\nu}$ and $\mathcal{R}_{K^*\nu\nu}$  above unity. A detailed updated numerical analysis
of these correlations should appear soon \cite{Buras:2023yyy}.

\subsection{Cabibbo Angle Anomaly and the Unitarity of the CKM\\ Matrix}
There was recently a large activity in connection with the so-called Cabibbo Angle Anomaly (CAA) which is nicely reviewed in  \cite{Crivellin:2022ctt}.
 There exist tensions between
different determinations of the elements $\vus$ and $\vud$ in the CKM matrix from different decays. For instance, the determination of $\vud$ from superallowed beta decays and of $\vus$ from kaon decays imply a violation of the first row
unitarity in the CKM matrix. There exist also tensions between determinations of $\vus$ from leptonic $K_{\mu2}$ and semileptonic $K_{l3}$ kaon decays.
In the end a $3\,\sigma$ deficit in the CKM unitarity
 relation corresponding to the first row of this matrix and less significant for  the first column exist:
 \be\label{column}
 \vud^2+\vus^2+\vub^2=0.9985(5),\qquad \vud^2+\vcd^2+\vtd^2=0.9970(18)\,.
 \ee
 As reviewed in \cite{Crivellin:2022ctt} numerous authors made efforts
 in the literature to find the origin of this anomaly and to remove it
 with the help of NP models containing $W^\prime$, vector-like
 leptons, vector-like quarks, scalars, $Z^\prime$ gauge boson and leptoquarks.

 Here we want to emphasize the following points:
 \begin{itemize}
 \item
   In the {\em absence} of new quarks, that must be vector-like, CKM unitarity
   {\em cannot} be violated. The violation in this case is only apparent due to possible contributions of bosons like $Z^\prime$ to decays used in the
   determination of 
   $\vud$ and $\vus$ or due to hadronic uncertainties or wrong measurements.
   Otherwise GIM mechanism would fail and moreover at one-loop level, as illustrated below,  gauge dependences would be present. 
 \item
   In the {\em presence} of vector-like quarks CKM unitarity {\em can} be violated with the CKM matrix being submatrix of a unitary matrix involving SM quarks and vector-like quarks. Adding then diagrams with quark and vector-like quark
   exchanges the GIM mechanism is  recovered.
 \end{itemize}

 Let us illustrate this problem repeating the 1974 Gaillard-Lee
 calculation \cite{Gaillard:1974hs} of the  $K_L-K_S$ mass difference in the SM
 that was performed in the Feynman gauge.

The basic formula for the relevant Hamiltonian used to calculate $\Delta M_K$ in the SM 
that allows to reach our goal can be found in equation (6.67) in \cite{Buras:2020xsm}.
It reads
\be\label{Htot1}
\boxed{\mathcal{H}^{ij}_{\rm eff}=\frac{G_F^2}{16\pi^2}M_W^2 \lambda_i \lambda_j \left[(1+\frac{x_ix_j}{4})\,T(x_i,x_j) -2x_ix_j\tilde T(x_i,x_j)\right] (\bar s d)_{V-A}(\bar s d)_{V-A}\,,}  ~~~~~~~~~~~~~~~~~~~~~~~~~~~
\ee
with
\be
\lambda_i =V_{is}^*V_{id},\qquad  x_i=\frac{m_i^2}{M^2_W}, \qquad i=u,c,t,
\ee
where $V_{ij}$ are the elements of the CKM matrix.
The functions $T(x_i,x_j)$ and $\tilde T(x_i,x_j)$ are given as follows
\begin{align}\label{HWW}
T(x_i,x_j)&=& \left[\frac{x^2_i\log x_i}{(1-x_i)^2(x_i-x_j)}+
  \frac{x^2_j\log x_j}{(1-x_j)^2(x_j-x_i)}+\frac{1}{(1-x_i)(1-x_j)}\right]\,,
~~~~~~~~~~~~~~~~~~~~~~~~~~~~~~~~~~~
\\
T(x_i,x_i)&=&\left[\frac{2x_i\log x_i}{(1-x_i)^3}+
\frac{1+x_i}{(1-x_i)^2}\right]\,,~~~~~~~~~~~~~~~~~~~~~~~~~~~~~~~~~~~~~~~~~~~~~~~~~~~~~~~~~~~~~~~~~~~~~~~~~~~~~~~~~~~
\\
\tilde T(x_i,x_j)&=& \left[\frac{x_i\log x_i}{(1-x_i)^2(x_i-x_j)}+
\frac{x_j\log x_j}{(1-x_j)^2(x_j-x_i)}+\frac{1}{(1-x_i)(1-x_j)}\right]\,,~~~~~~~~~~~~~~~~~~~~~~~~~~~~~~~~~~~~\\
\tilde T(x_i,x_i)&=&\left[\frac{(1+x_i)\log x_i}{(1-x_i)^3}+
\frac{2}{(1-x_i)^2}\right]\,.~~~~~~~~~~~~~~~~~~~~~~~~~~~~~~~~~~~~~~~~~~~~~~~~~~~~~~~~~~~~~~~~~~~~~~~~~~~~~~
\end{align}
Detailed derivation of these formulae can be found in Section 6.2 of \cite{Buras:2020xsm}. 

Summing 
over the internal up-quarks and using the relation
\be\label{UTR}
\lambda_u+\lambda_c+\lambda_t=0\,,
\ee
that follows from the unitarity of the CKM matrix, we find the well known
SM expressions. For our purposes it is sufficient to keep only the term
involving $\lambda^2_c$ because this term dominates the SM contribution
to $\Delta M_K$. Using the standard formulae one finds then
\be\label{DMK1}
\Delta M_K=\frac{G_F^2}{6\pi^2}\hat B_K F_K^2M_W^2\eta_1 \lambda_c^2 x_c\,,
\ee
where the mass of the up-quark has been set to zero. $\eta_1$ is known at NNLO  in QCD and is subject to a large uncertainty but this
is not relevant for our main argument. Other factors are also well known.

On the other hand, not using (\ref{UTR}) and keeping only mass-independent terms  and $\ord(m^2_c/M_W^2)$ terms we find
\be\label{AJB}
\Delta M_K=2.52 \, T(x_c)\, 10^{-10}\GeV
\ee
where
\be\label{TXC}
T(x_c)= (\lambda_u+\lambda_c)^2+\lambda^2_c F_1(x_c)+2\lambda_u\lambda_c F_2(x_c)
\ee
with
\be
F_1(x_c)=2 x_c \log x_c + 3 x_c,\qquad F_2(x_c)= x_c \log x_c +  x_c.
\ee

Going back to 1974 and using the unitarity relation valid for
two quark generations
\be\label{UTR2}
\lambda_u+\lambda_c=0\,,
\ee
instead of (\ref{UTR}), we find 
$T(x_c)=\lambda_c^2 x_c $ and consequently (\ref{DMK1}).
However, if this relation is violated because of the CAA, the first mass independent contribution
and in particular the logarithmic terms, that do not cancel each other, have a large impact on the final result. Indeed
we find
\begin{align}\label{FF}
  F_1(x_c)&=&-41.4 \cdot 10^{-4} +7.5 \cdot 10^{-4}=-33.9\cdot10^{-4} \,,~~~~~~~~~~~~~~~~~~~~~~~~~~~~~~~~~~~~~~~~~~~~\\
  F_2(x_c)&=& -20.7 \cdot 10^{-4} +2.5\cdot 10^{-4}= -18.2\cdot10^{-4}\,.~~~~~~~~~~~~~~~~~~~~~~~~~~~~~~~~~~~~~~~~~~~~
\end{align}
We observe that the unitarity relation (\ref{UTR2}) is very important in
cancelling the logarithmic terms that are roughly by an order of magnitude
larger than  the SM result for $\Delta M_K$. But also  the modification due to the first term in (\ref{TXC}), that is not mass suppressed, is important.

However, these are minor problems in comparison with the fact that these results are gauge dependent and the ones just presented correspond to the Feynman gauge with the
gauge parameter $\xi=1$. Concentrating on the massless limit and repeating
the calculation in an arbitrary covariant gauge we find
\be
T(0)=  (\lambda_u+\lambda_c)^2\,\left[1-2\left(1+\frac{\xi}{1-\xi}\ln\xi\right)
  +\frac{1}{4}\left(1+\xi +\frac{2\xi}{1-\xi}\ln\xi\right)\right].
\ee
We note that for $\xi=1$ we reproduce (\ref{TXC}) for $x_c=0$ and that in the unitary gauge for which $\xi\to\infty$, $T(0)$ diverges. It vanishes only
if the unitarity relation in  (\ref{UTR2})is imposed, the crucial property
in the GIM mechanism.

Let us next return to the SM and consider three generations and the
$B^0_{s,d}-\bar B^0_{s,d}$ mass differences $\Delta M_{s,d}$. 
In this case the well known unitarity relations apply
\be\label{CKM}
\lambda_u+\lambda_c+\lambda_t= 0, \qquad \lambda_i=V^*_{ib}V_{iq},\qquad q=s,d\,.
\ee
If these relations are violated the results for $\Delta M_{s,d}$ are gauge dependent and divergent in the unitary gauge. In fact this has been demonstrated already in 2004 \cite{Buras:2004kq}  in the context  of the analysis in the littlest Higgs Model  in which CKM unitarity is violated due to the presence of a heavy
vector-like quark $T$ that mixes with the ordinary quarks.

In order to obtain gauge independent and finite results the CKM unitarity relation in (\ref{CKM}) should be replaced by the new one that reads
\be\label{GCKM}
\hat\lambda_u+\hat\lambda_c+\hat\lambda_t+\hat\lambda_T = 0 \qquad \hat\lambda_i=\hatV^*_{ib}\hat V_{iq}, \qquad q=s,d\,,
\ee
with $\hat V_{ij}$ containing corrections from the mixing with $T$ in question.
Explicit formulae for $\hat V_{ij}$ are given in \cite{Buras:2004kq}.
To my knowledge this was the first paper which explicitly demonstrated
 that only
 after the imposition of this unitarity relation
 the divergences in the unitary gauge could be removed.

 But now comes an important point made in \cite{Buras:2004kq}.
 In order for the relation in (\ref{GCKM}) to be effective it is crucial
 to include box diagrams involving both SM quarks and the vector-like quark $T$
 in addition to the usual box-diagrams involving ordinary quarks only
 and the ones that involve $T$ only. This is evident from the expressions above
 and applies to all VLQ models. More recent calculations of this type
 in the context of vector-like models can be found in
 \cite{Belfatto:2019swo,Belfatto:2021jhf,Botella:2021uxz,Crivellin:2022rhw}.
 Most recent extensive review of the vector-like quarks can be found in \cite{Alves:2023ufm} and in Chapter 16.3 of \cite{Buras:2020xsm}.

\boldmath
\section{Dirac vs. Majorana in Rare $K$ and $B$ Decays}
\unboldmath
 \subsection{Preliminaries}
 Most of the analyses of $K\to\pi\nu\bar\nu$ and $B\to K(K^*)\nu\bar\nu$ decays in the literature assume that neutrinos are Dirac particles and consequently
 the conservation of the lepton number (LNC). This implies, as we have seen in
Section~\ref{Zprime},  certain patterns
  of the correlations between $\klpn$ and $\kpn$ branching ratios that  depend on the NP  scenario   considered. Analogous
  correlations are found between branching ratios for  $B\to K\nu\bar\nu$ and $B\to K^*\nu\bar\nu$  decays \cite{Altmannshofer:2009ma,Buras:2014fpa}. In certain models correlations between all four decays are present.
  The question then arises how these correlations would be modified if
  neutrinos were Majorana particles.

  Recently, the role of lepton number violating (LNV) scalar operators and specifically of Majorana neutrinos in the four decays in question 
has been addressed in a number of papers \cite{Li:2019fhz,Deppisch:2020oyx,Deppisch:2020zrd,Felkl:2021uxi}. The main goal of these papers was to derive a number of useful formulae for the new contributions to $\kpn$ and $\klpn$ that
are represented by a single dimension-7 operator within the SM effective field
theory (SMEFT). Having these formulae it was possible 
to derive the bounds on NP scale setting the relevant couplings to unity. Moreover in \cite{Deppisch:2020oyx,Deppisch:2020zrd} kinematic
distributions for $\kpn$ and $\klpn$ have been demonstrated to be sensitive
probes of LNV. A comprehensive survey of dimension-7 SMEFT operators in the context of LNV has been recently presented in \cite{Fridell:2023rtr}.

Here I would like to report briefly  preliminary results from a recent work  \cite{Buras:2023xxx} in collaboration with Julia Harz  that developed new efficient strategies that would allow the experimentalists of   NA62, KOTO and Belle II collaborations  to tell us one day whether the footprints of Majorana   neutrinos are present in their data. The major role in these strategies   play the distributions in $s$, the invariant mass-squared of the neutrino system, that allow the separation of vector current (Dirac) contributions to all these decays from the scalar (Majorana) ones. Here we summarize the
main results of this analysis.
\boldmath
\subsection{$\kpn$ and $\klpn$}\label{sec:2M}
\unboldmath
Let us denote the Wilson coefficients representing vector and scalar operators
as follows
\be
C_V=|C_{\rm SM}|e^{i\phi_{\rm SM}}+|C^\text{NP}_V|e^{i\phi_V},\qquad
C_S=|C_S|e^{i\phi_S}\,.
\ee

\begin{figure}[th]
\centering
\includegraphics[width=\textwidth]{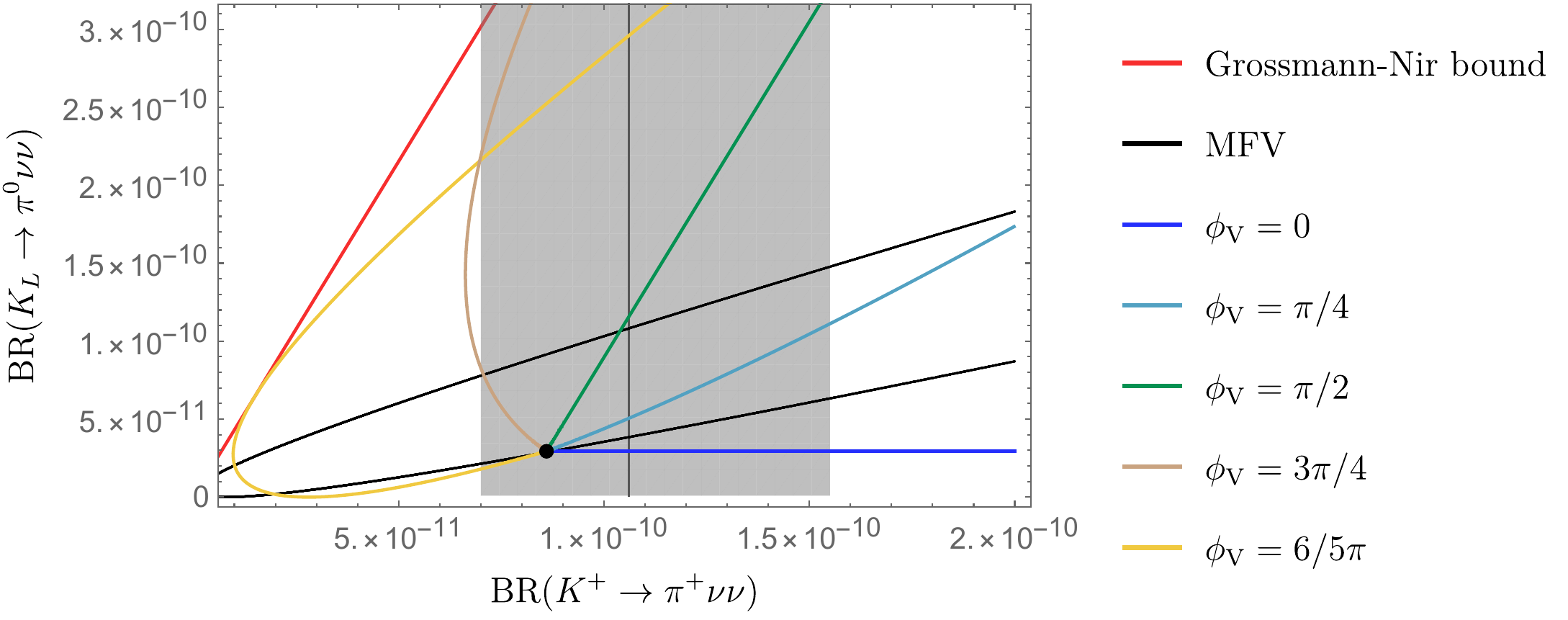}\\
\includegraphics[width=\textwidth]{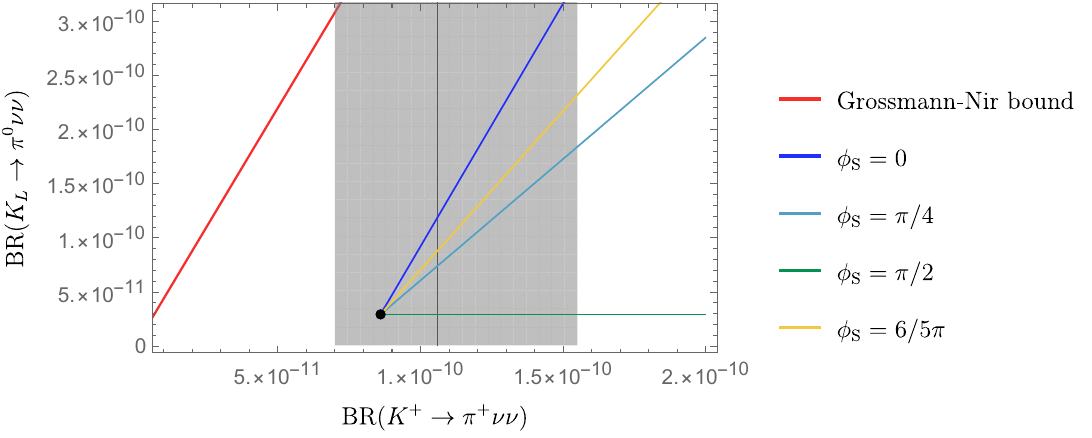}
\caption{$\mathcal{B}(\kpn)-\mathcal{B}(\klpn)$-plane for the case of pure NP
  vector contribution with $C_S=0$ (top) and pure NP scalar contributions with $C_V=0$ (bottom). 
    The red line indicates the Grossmann-Nir bound. The SM contribution  is represented by a dark point. {Additionally,}  we allow for  NP contributions with  $\phi_V$  and $\phi_S$ fixed to specific values and varying $C_V$ and $C_S$, respectively. The grey region represents the present experimental $1\sigma$ range. From \cite{Buras:2023xxx}.}
\label{fig:KaonVectorScalar}
\end{figure}

The outcome of the numerical analysis in our paper  is presented
  in Fig.~\ref{fig:KaonVectorScalar}. In the upper plot we show the action
  of a pure NP   {\em vector} contribution with $C_S=0$ in the 
  $\mathcal{B}(\kpn)-\mathcal{B}(\klpn)$-plane. In the lower plot
  we show the corresponding {impact} of a pure NP   {\em scalar} contribution with $C_V=0$. In both cases the SM contribution is represented
  by a dark point, the central experimental value of $\mathcal{B}(\kpn)$ by
  a vertical black line and the GN bound by a red line.

  Let us summarize the main observations concerning the above relations and the resulting plots in     Fig.~\ref{fig:KaonVectorScalar}.
       \begin{itemize}
       \item
         The cases $\phi_V=\pi/2$ ({{\em green}})  and
         $\phi_V=0$ ({{\em blue}}) are the two branches pointed out in 
         \cite{Blanke:2009pq} and discussed by us already in
         Section~\ref{MBABK}. For $0<\phi_V<\pi/2$ the correlation between
       the two branching
       ratios takes place on lines between the two branches found above with
       the slopes {\em increasing} with increasing $\phi_V$.
       The {black solid} line in the upper plot represents the case of Minimal
         Flavour Violation (MFV). 
\item
      The novel feature is the impact of pure {\em scalar} contributions shown in the lower part of Fig.~\ref{fig:KaonVectorScalar}. While for $\phi_V=0$ {the} NP vector contribution to $\mathcal{B}(\klpn)$ vanishes, in the scalar case it is maximal for $\phi_S=0$. For a non-vanishing $\phi_S$ the correlation between the two       branching ratios proceeds on a line which is not parallel to the
      GN line but which has a slope which {\em decreases}, as opposed to the vector case, with increasing       $\phi_S$ for  $0\le\phi_S\le 90^\circ$.
\item
      When comparing the two  plots in  Fig.~\ref{fig:KaonVectorScalar},
      it becomes clear that a scalar contribution can only increase the branching ratios, while a vector contribution can also decrease them with respect to the SM model expectation. The scalar contribution is solely confined between the green ($\phi_S=\pi/2$) and blue ($\phi_S= 0$) lines, and does not extend to the parameter space where for example the yellow curve ($\phi_V= 6/5\pi$) in the
      upper part in Fig.~\ref{fig:KaonVectorScalar} is found. Hence a deviation from the SM to lower values would point towards a NP vector contribution, while excluding a NP contribution from a LNV scalar current \emph{only}.
    \item
    When allowing for all four non-vanishing  NP parameters
    \be
      |C_{V}|, \quad \phi_{V}, \quad |C_{S}|, \quad \phi_{S}
      \ee
    at the same time, all the parameter space below the Grossmann-Nir bound (red solid line) in the $\mathcal{B}(\kpn)-\mathcal{B}(\klpn)$-plane is possible, while a pure scalar NP contribution is confined to the area between the blue and green lines in the lower part  in  Fig.~\ref{fig:KaonVectorScalar}. Hence when measuring lower $\kpn$ and $\klpn$ branching ratios than expected in the SM, a scalar current can only be present with an additional vector contribution.
          \end{itemize}         

These results demonstrate very clearly the different impact of vector and scalar contributions on the $\mathcal{B}(\kpn)-\mathcal{B}(\klpn)$-plane. But they  also make clear that  the branching ratios alone will not allow us to identify {a possible underlying vector or scalar current.}  Fortunately,
       as pointed out already in \cite{Li:2019fhz,Deppisch:2020oyx,Deppisch:2020zrd} this is possible with the help of kinematic distributions for $\kpn$ and $\klpn$, in particular with the help of    
       the distributions in $s$, the invariant mass-squared of the neutrino system. They indeed allow the separation of vector current contributions from scalar ones   without specifying a NP model. A first look in this direction
       has been made in \cite{Li:2019fhz,Deppisch:2020oyx,Deppisch:2020zrd}.
Here, I summarize the results obtained in collaboration with Julia Harz.
 \begin{figure}
\centering
\includegraphics[width=0.8\textwidth]{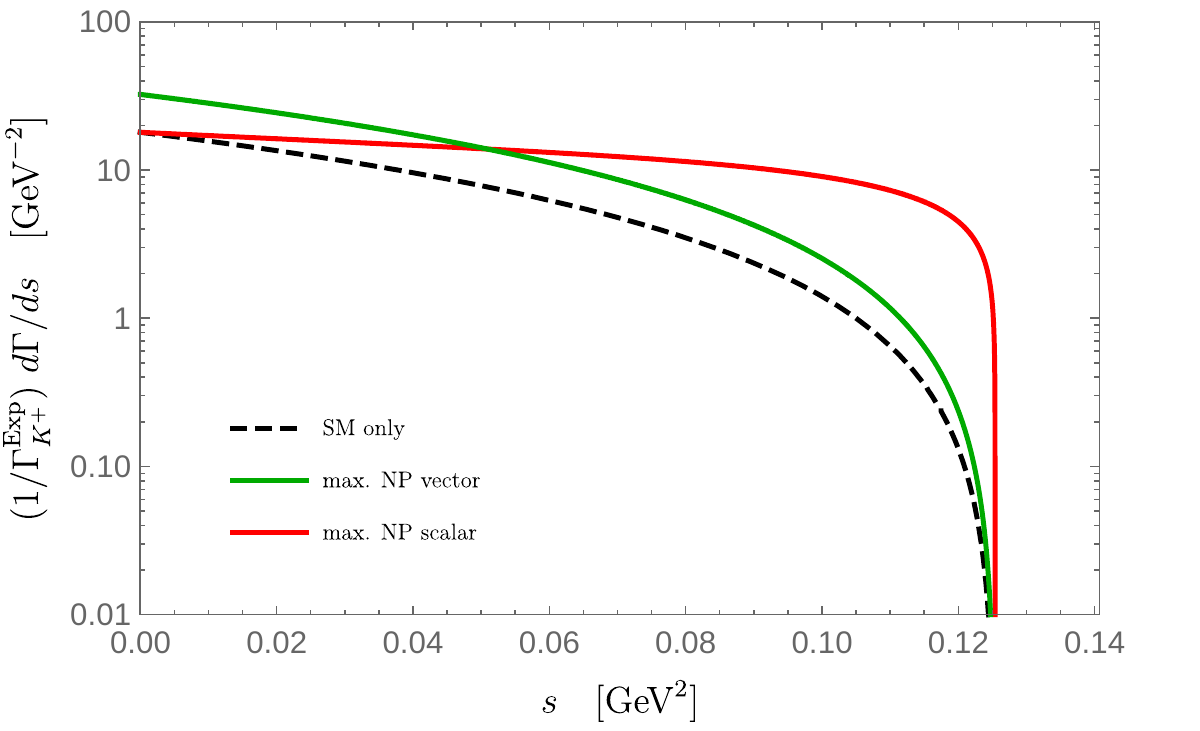}	
\vspace{0.5cm}
\includegraphics[width=0.8\textwidth]{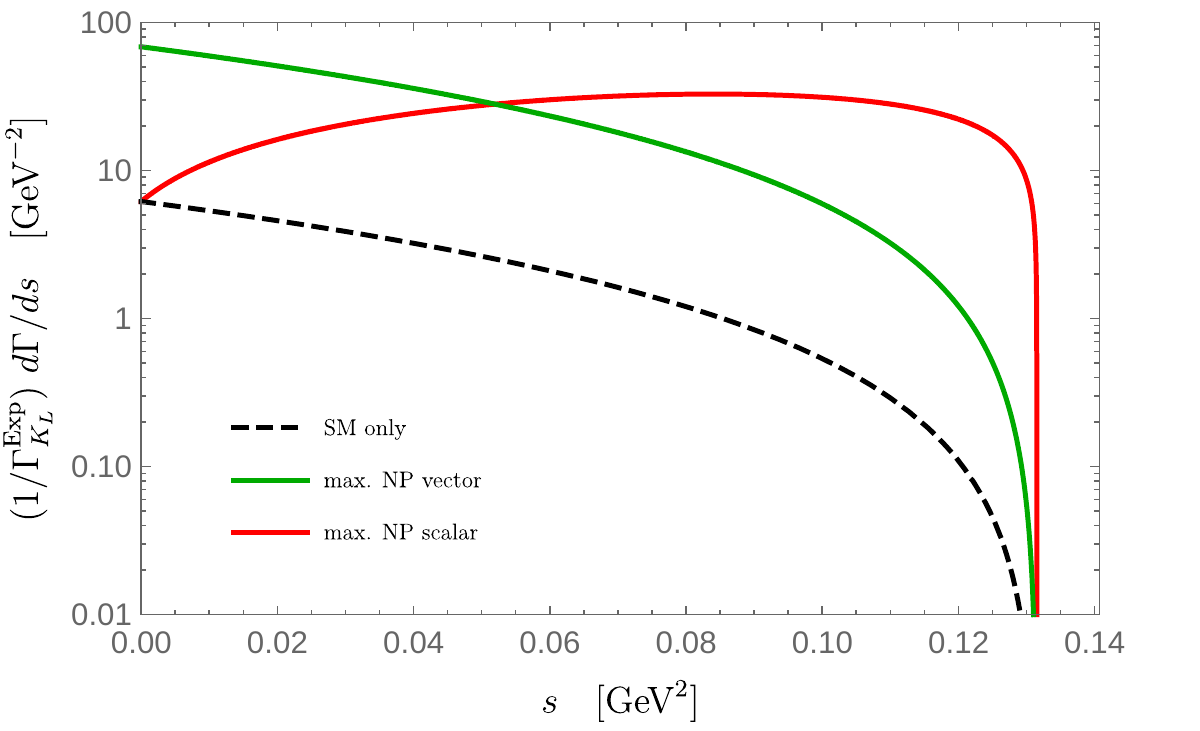}	
\caption{{{Differential distribution of $\Gamma(\kpn)/ds$ (upper plot) and $\Gamma(\klpn)/ds$ (lower plot) normalized to the total experimental decay width $\Gamma_{K^+}^{\mathrm{Exp}}$ and $\Gamma_{K_L}^{\mathrm{Exp}}$, respectively. The SM contribution only is depicted in black dashed. For the red (green) line we have assumed an additional scalar (vector) contribution besides the SM contribution, leading to a NP signal around the current experimental upper limit $\mathcal{B}(\kpn)=1.55 \times 10^{-10}$.  For the vector contribution a phase of $\phi_V=\pi/2$ was chosen. From \cite{Buras:2023xxx}.}}}
\label{fig:DistKLp}
\end{figure}

In Fig.~\ref{fig:DistKLp}, we depict the differential distribution of $\Gamma(\kpn)/ds$ (upper plot) and $\Gamma(\klpn)/ds$ (lower plot) normalized to the total experimental decay width $\Gamma_{K^+}^{\mathrm{Exp}}$ and $\Gamma_{K_L}^{\mathrm{Exp}}$, respectively. We show the expected SM contribution by the black
dashed line. For the red (green) line we have assumed an additional scalar (vector) contribution besides the SM contribution, leading to a NP signal around the current experimental upper limit $\mathcal{B}(\kpn)=1.55 \times 10^{-10}$.  For the vector contribution a phase of $\phi_V=\pi/2$ was chosen.
For the scalar contribution (red) we fix $\phi_S=0$, leading to the maximal contribution to $\mathcal{B}(\klpn)$. The following observations can be made:
\begin{itemize}
 \item While the vector contribution follows the shape of the SM contribution, the scalar contribution features a distinct distribution, clearly different from the vector one, in particular for larger $s$. Even within the current possible range for NP  (limited by the current experimental upper limit $\mathcal{B}(\kpn)=1.55 \times 10^{-10}$), the difference is strong enough to be visible in the distribution.
 \item Interestingly, we find that the scalar differential distribution $\Gamma(\kpn)/ds$ turns out to be independent of the scalar phase $\phi_S$, while
   $\Gamma(\klpn)/ds$ is. Hence, for different  $\phi_S$, $\Gamma(\kpn)/ds$ does not change, while $\Gamma(\klpn)/ds$ does. While it is maximal for $\phi_S=0$, it follows the SM line for $\phi_S=\pi/2$. Therefore, a combined analysis of the differential distributions for $\Gamma(\kpn)$ and $\Gamma(\klpn)$ is very powerful. In case $\Gamma(\kpn)$ would give a hint towards a new scalar contribution, a comparison with $\Gamma(\klpn)$ could tell us if it features a non-zero scalar phase $\phi_S$. 
\end{itemize}

\boldmath
\subsection{$B\to K\nu\bar\nu$ and $B\to K^*\nu\bar\nu$}\label{sec:2Mstar}
\unboldmath
While the search for the footprints of Majorana neutrinos can also be done in this case with the help of $s$-distributions, it can also be performed with integrated quantities.
To this end the following two model-independent relations pointed out in  \cite{Buras:2014fpa} can be used. The first one
is
\begin{equation}
  \boxed{\langle F_L \rangle = \langle F_L^\text{SM} \rangle
  \left(\frac{(\kappa_\eta-2)\mathcal{R}_{K\nu\nu}+4\,\mathcal{R}_{K^*\nu\nu}}{(\kappa_\eta+2)\mathcal{R}_{K^*\nu\nu}}\right)r_1^{\rm LNV}
\,}
\label{eq:FLtest}
\end{equation}
with $\mathcal{R}_{K\nu\nu}$ and  $\mathcal{R}_{K^*\nu\nu}$  defined in (\ref{RKRK*}). 
It is given here in the integrated form, but
in principle it can be tested experimentally also on a bin-by-bin basis.
It is valid in the presence of LFV and lepton flavour universality violation.
As pointed out by us this relation is violated
in the presence of lepton number violating (LNV) contributions which we indicated by introducing a new
parameter $r_1^{\rm LNV}$ that equals unity in the case of lepton number conserving (LNC) contributions but differs from
it in the case of LNV contributions, that is scalar contributions in our
case. That this relation is violated in the presence of scalar contributions follows simply from the fact
that  the three observables in (\ref{eq:epseta-R1}) depend only on two parameters $\epsilon$ and $\eta$. The presence of scalar contributions introduces 
new parameters and the relation in (\ref{eq:FLtest}) with  $r_1^{\rm LNV}=1$ does not apply any longer.

A similar relation can be obtained for the modification of the inclusive $B\to X_s\nu\bar\nu$ branching ratio  \cite{Buras:2014fpa},
\begin{equation}
\boxed{\mathcal{B}(B\to X_s\nu\bar\nu)
=
\mathcal{B}(B\to X_s\nu\bar\nu)_\text{SM}\left(
\frac{\kappa_\eta \mathcal{R}_{K\nu\nu}+2\,\mathcal{R}_{K^*\nu\nu}+0.09(\mathcal{R}_{K^*\nu\nu}-\mathcal{R}_{K\nu\nu})}{\kappa_\eta+2}
\right)r_2^{\rm LNV}\,.}
\label{eq:Xstest}
\end{equation}
The new parameter $r_2^{\rm LNV}$ introduced here again allows to describe the
effects of LNV contributions and equals unity in the LNC case.

We emphasize again that the relations (\ref{eq:FLtest}) and (\ref{eq:Xstest}) hold even in the case of lepton flavour non-universality and lepton flavour violation. Consequently, a violation of either of them unambiguously signals either
the presence of particles other than neutrinos in the final state (as discussed e.g.\ in \cite{Altmannshofer:2009ma,Schmidt-Hoberg:2013hba}) or Majorana neutrinos. Detailed new analysis should be presented soon \cite{Buras:2023xxx}

\boldmath
\section{$\Delta I=1/2$ rule and $\epe$}
\unboldmath
\boldmath
\subsection{QCD dynamics and the $\Delta I=1/2$ rule\label{sec:2b}}
\unboldmath
One of the puzzles of the 1950s was a large disparity between the measured 
values of the real parts of the isospin amplitudes $A_0$ and $A_2$ in $K\to\pi\pi$ decays, which on the basis of usual isospin 
considerations were expected to be of the same order.

In 2023 we know the 
experimental values of the real parts of these amplitudes very precisely 
\cite{Tanabashi:2018oca}
\bea\label{N1}
{\rm Re}A_0= 27.04(1)\times 10^{-8}~\GeV, \nn\\
\quad {\rm Re}A_2= 1.210(2)   \times 10^{-8}~\GeV.
\eea
As ${\rm Re}A_2$ is dominated by $\Delta I=3/2$ transitions but 
${\rm Re}A_0$ receives contributions also from $\Delta I=1/2$ transitions, 
the latter transitions dominate ${\rm Re}A_0$  which expresses
the so-called $\Delta I=1/2$ rule \cite{GellMann:1955jx,GellMann:1957wh}
\be\label{N1a}
{R}=\frac{{\rm Re}A_0}{{\rm Re}A_2}=22.35.
\ee

Soon after the discovery of asymptotic freedom  in 1973 Altarelli and Maiani \cite{Altarelli:1974exa} and Gaillard and Lee \cite{Gaillard:1974nj} made a first unsuccessful attempt to explain this  huge enhancement 
through short distance QCD effects. As we already reported, 
the precision of the calculation of the short distance contributionss increased  considerably in the last fifty years since this
first pioneering calculation. The basic QCD
dynamics behind this rule - contained in the hadronic matrix elements
of current-current operators - has been identified analytically first in 1986 in the framework of the Dual QCD in \cite{Bardeen:1986vz} with some improvements
in 2014  \cite{Buras:2014maa}. This has been  confirmed more than 30 years later
by the RBC-UKQCD collaboration \cite{RBC:2020kdj} although the modest accuracy of both approaches still allows for some NP contributions. See
\cite{Buras:2022cyc} for the most recent summary. Despite this summary it is appropriate to describe here briefly  the present situation of this important rule that is governed by QCD dynamics.

Let us then start 
by evaluating the 
simple $W^\pm$ boson exchange between the relevant quarks which after integrating out $W^\pm$ 
generates the current-current operator $Q_2$:
\begin{equation}\label{O1EK} 
Q_2 = (\bar s\gamma_\mu(1-\gamma_5)u)\;(\bar u\gamma^\mu(1-\gamma_5)d)~.
\end{equation}
With only $Q_2$ contributing we have 
\be
 {\rm Re}A_{0,2}=\frac{G_F}{\sqrt{2}}V_{ud}V_{us}^*\langle Q_2\rangle_{0,2}\,.
\ee
Calculating  the matrix elements $\langle Q_2\rangle_{0,2}$ in the strict large $N$ limit, which corresponds to factorization of matrix elements of $Q_2$ into 
the product of matrix elements of currents, we find 
\be\label{Q2N}
\langle Q_2\rangle_{0}=\sqrt{2}\langle Q_2\rangle_{2}=\frac{2}{3}\,F_\pi(m_K^2-m_\pi^2),
\ee
with $F_\pi$ being pion weak decay constant and consequently
\be\label{LO}
{\rm Re}A_0=3.59\times 10^{-8}\GeV ,\qquad  
{\rm Re}A_2= 2.54\times 10^{-8}\GeV~, \qquad  R=\sqrt{2},
\ee
in plain disagreement with the data in (\ref{N1}) and (\ref{N1a}). 

It should be emphasized that the explanation of the  missing enhancement factor of $15.8$ in ${R}$ through some dynamics must simultaneously give the correct values for ${\rm Re}A_0$ and  ${\rm Re}A_2$. 
This means that this dynamics should suppress  ${\rm Re}A_2$ by a factor of $2.1$, not more, and enhance ${\rm Re}A_0$ by a factor of $7.5$. This tells us 
that while the suppression of  ${\rm Re}A_2$  is an important ingredient in 
the $\Delta I=1/2$ rule, it is not the main origin of this rule.
 It is the enhancement of  ${\rm Re}A_0$  as
 already emphasized in \cite{Shifman:1975tn}.  However, in contrast to this paper, the 
 current-current operators, like $Q_2$, are responsible dominantly for this rule and not QCD penguins. This was
 pointed out first in 1986  \cite{Bardeen:1986vz} and  demonstrated in the context of the Dual QCD approach.
 An update and improvements over the 1986 analysis appeared in 2014 \cite{Buras:2014maa} with the result
\be\label{NLO+M+P}
{R}\approx 16.0\pm 1.5~,  \qquad {\rm DQCD~~(1986,2014)},
\ee
that is one order of magnitude enhancement over the result in (\ref{LO})
without QCD up to confinement of quarks in mesons.
The missing piece could come from final state interactions as stressed in particular 
by ChPT experts \cite{Pallante:2000hk}. Also $1/N^2$ corrections could  change this result but are unknown.

Meanwhile the RBC-UKQCD LQCD collaboration confirmed in 2012 the 1986 DQCD finding
that current-current operators dominate the $\Delta I=1/2$ rule. But
the results from the series of their three papers show how difficult
these calculations on the lattice are:  $R=12.0\pm1.7$~\cite{Boyle:2012ys},
$R=31.0\pm 11.1$~\cite{Blum:2015ywa} and finally  \cite{RBC:2020kdj}
\be\label{EB}
\frac{\RE A_0}{\RE A_2}=19.9(2.3)(4.4), \qquad {\rm RBC-UKQCD~~(2020)~}
\ee
 that is consistent with the DQCD value and  in  agreement
with the experimental value $22.4$.
While the RBC-UKQCD result is closer to the data than the DQCD one,
the dynamics behind this rule, except for the statement that it is QCD,
has not been provided by these authors. To this end it is necessary
to switch off QCD interactions which can be done in the large $N$ limit in
DQCD but it seems to be impossible or very difficult on the lattice.

The anatomy of QCD dynamics as seen within the DQCD approach has been
presented in \cite{Bardeen:1986vz,Buras:2014maa} and in particular
in Section 7.2.3 of \cite{Buras:2020xsm}. Here we just present an express
view of this dynamics.

Starting with the values in (\ref{LO}), the first step is to include
the
short-distance  RG-evolution of Wilson Coefficients (WCs) from scales ${\cal O}(M_W)$
down to scales in the ballpark of $1~\GeV$. This is the step made already 
in the pioneering 1974 calculations in \cite{Altarelli:1974exa,Gaillard:1974nj} except that they were done 
 at LO in the  RG-improved perturbation theory 
 and now can be done at the NLO level.
These 1974 papers have shown that the short distance QCD effects enhance ${\rm Re}A_0$ and suppress ${\rm Re}A_2$. However, the inclusion of NLO QCD corrections to WCs of $Q_2$ and $Q_1$ operators \cite{Altarelli:1980fi,Buras:1989xd}
made it clear, as stressed in particular in \cite{Buras:1989xd}, that the 
 $K\to\pi\pi$ amplitudes without the proper calculation of hadronic matrix 
elements of $Q_i$ are both scale and renormalization-scheme dependent.
Moreover, further enhancement 
of ${\rm Re}A_0$ and further suppression of ${\rm Re}A_2$ are needed in order 
to be able to understand the $\Delta I=1/2$ rule.

This brings us to the second step first performed in 1986 in \cite{Bardeen:1986vz}
within the DQCD approach.
Namely, the RG-evolution down to the scales $\ord(1\GeV)$ is continued as a short but fast {\em meson evolution} down to zero momentum scales at which the factorization of hadronic matrix elements is at work and one can in no
time calculate the hadronic matrix elements in terms of meson masses and weak decay constants as seen in (\ref{Q2N}).
Equivalently, starting with factorizable hadronic matrix elements
of current-current operators
at $\mu\approx 0$ 
and evolving them to $\mu=\ord(1\GeV)$ at which the WCs are evaluated  one is able 
to calculate the matrix elements of these  operators at $\mu=\ord(1\GeV)$ and properly combine them with their WCs evaluated at this scale.
The final step is the inclusion of QCD penguin operators that provide
an additional enhancement of $A_0$ by roughly $10\%$ without changing $A_2$.

In \cite{Bardeen:1986vz} only the pseudoscalar meson contributions to
meson evolution have been included and the {\em quark evolution}, RG evolution above $\mu=\ord(1\GeV)$, has been performed at LO. The improvements in 2014
\cite{Buras:2014maa} were the inclusion of vector meson contributions
to the meson evolution and the NLO corrections to quark evolution. These
improvements practically removed scale and renormalization-scheme dependences
and brought  the theory closer to data.

Based on DQCD and RBC-UKQCD results we conclude 
that the QCD dynamics is dominantly responsible for the $\Delta I=1/2$ rule.
However, in view of large uncertainties in both DQCD and RBC-UKQCD
results, NP contributions at the level of $15\%$ could still be present.
See \cite{Buras:2014sba} to find out what this NP could be.

Finally other authors  suggested different explanations of the $\Delta I=1/2$
rule within QCD that were published dominantly in the 1990s and their list can be
found in \cite{Buras:2020xsm}. But in my view the DQCD picture
of what is going on is more beautiful and transparent as asymptotic
freedom and related non-factorizable QCD interactions are primarily responsible for this rule. 
 It is simply the
 {\em quark evolution} from $M_W$ down to scale $\ord(1\GeV)$ as analysed first     by  Altarelli and Maiani \cite{Altarelli:1974exa} and Gaillard and Lee \cite{Gaillard:1974nj}, followed by the  {\em meson evolution} \cite{Bardeen:1986vz,Buras:2014maa} down to very low scales at which  QCD becomes  a theory of weakly interacting mesons and  a free theory of mesons in the strict large $N$ limit,
 a point made by 'tHooft and Witten in the 1970s.

\boldmath
    \subsection{QCD Dynamics and the Ratio $\epe$}
    \label{subsubsec:eps}
    \unboldmath
    While the parameter $\varepsilon\equiv\varepsilon_K$ measures the indirect CP-violation in $K_L\to\pi\pi$ decays, that is originating in the $K^0-\bar K^0$ mixing, the parameter $\varepsilon^\prime$ describes the direct CP violation, that is  in the decay itself.
   
    The story of $\epe$ both in the theory and experiment
    has been described in detail in \cite{Buras:2020wyv}. On the experimental
    side the chapter on $\epe$ seems to be closed for the near future. After heroic efforts, lasting
    15 years, the experimental world average of $\epe$ 
from NA48 \cite{Batley:2002gn} and KTeV
\cite{AlaviHarati:2002ye,Abouzaid:2010ny} collaborations reads
\be\label{EXP}
(\epe)_\text{exp}=(16.6\pm 2.3)\times 10^{-4} \,.
\ee
    On the theoretical side the first calculation of $\epe$ that included RG QCD effects to QCD penguin (QCDP) contributions is due to Gilman and Wise \cite{Gilman:1978wm} who - following Shifman, Vainshtein and Zakharov \cite{Shifman:1975tn} - 
assumed that {the} $\Delta I=1/2$ rule is explained by QCDP. Using the required values of
{the} QCDP  matrix elements for the explanation of this rule, they predicted $\epe$ to be in the ballpark of $5\times 10^{-2}$. During the 1980s this value decreased by roughly a factor
of 50 dominantly due to three effects:
\begin{itemize}
\item
  The first calculation of hadronic matrix elements
of QCDP operators in QCD,  carried out  in 
the framework of the DQCD \cite{Bardeen:1986vp,Bardeen:1986uz,Bardeen:1986vz}
in the strict large $N$ limit of colours, demonstrated, as already stated above,  that
QCDPs are not responsible for the $\Delta I=1/2$ rule and 
their hadronic matrix elements are much smaller than used in Gilman-Wise calculation.
\item
  The QCDP contribution to $\epe$ through isospin breaking in the
  quark masses \cite{Donoghue:1986nm,Buras:1987wc} {is significantly suppressed}.
\item
  The suppression of $\epe$ by electroweak penguin (EWP) contributions, that enter $\epe$ with the opposite sign to QCDP's contribution, is increased by the large top quark mass \cite{Flynn:1989iu,Buchalla:1989we}.
\end{itemize}

In the 1990s these calculations have been refined through NLO QCD calculations
to both QCDP and EWP contributions   by the  Munich and Rome teams:  \cite{Buras:1991jm,Buras:1992tc,Buras:1992zv,Buras:1993dy} and \cite{Ciuchini:1992tj,Ciuchini:1993vr}, respectively. In \cite{Buras:1999st} the NNLO QCD effects
on EWP contributions have been calculated. The NNLO QCD effects on QCDP
contributions are expected to be known in 2023. 

These NLO and NNLO QCD contributions
decreased various scale and renormalization-scheme uncertainties and suppressed $\epe$ within the SM further so that already in 2000 we knew that this ratio 
{should be of the order} 
of $1.0\times 10^{-3}$. Unfortunately 
even today the theorists  do not agree on whether the SM
agrees with the experimental value in (\ref{EXP}) or not.
The reason are different estimates of non-perturbative hadronic
QCD effects. This has been summarized recently in \cite{Buras:2022cyc}.
We recall only the main points below.

$\varepsilon^\prime$ is governed by 
the real and imaginary parts of the isospin amplitudes $A_0$ and $A_2$
 so that $\epe$ is given by  \cite{Buras:2015yba} 
\be\label{eprimea}
{\frac{\varepsilon'}{\varepsilon} = -\,\frac{\omega_+}{\sqrt{2}\,|\varepsilon|}
\left[\, \frac{{\IM} A_0}{{\RE}A_0}\,(1-\hat\Omega_{\rm eff}) -
\frac{1}{a}\frac{{\IM} A_2}{{\RE}A_2} \,\right],}
\ee
with $(\omega_+,a)$ and $\hat\Omega_{\rm eff}$ given in 2023
as follows
\be\label{OM+}
\omega_+ =a\,\frac{\RE A_2}{\RE A_0} =(4.53\pm0.02)\times 10^{-2}
\ee
with $a=1.017$ and
\be\label{omega9}
\hat\Omega_{\rm eff} = (29\pm 7)\times 10^{-2}\,.
\ee
Here $a$  and $\hat\Omega_{\rm eff}$ summarize  isospin breaking corrections and include  strong isospin
violation $(m_u\neq m_d)$, the correction to the isospin limit coming from
$\Delta I=5/2$ transitions and  electromagnetic corrections \cite{Cirigliano:2003nn,Cirigliano:2003gt,Bijnens:2004ai}. The most recent value for $\hat\Omega_{\rm eff}$ given above includes the nonet of pseudoscalar { meson}s and $\eta-\eta^\prime$
mixing  \cite{Buras:2020pjp}. If only the octet of pseudoscalar {meson}s is included
so that $\eta-\eta^\prime$ mixing does not enter, as presently done in ChPT,
one finds $\hat\Omega_{\rm eff}=(17\pm9)\, 10^{-2}$ \cite{Cirigliano:2019ani}, 
a value called $\hat\Omega_{\rm eff}^{(8)}$ here.
The inclusion of $\eta-\eta^\prime$ mixing { yields $\hat\Omega_{\rm eff}^{(9)}$ 
in (\ref{omega9}).
This contribution} is important,
a fact known already for 35 years \cite{Donoghue:1986nm,Buras:1987wc}.

In the SM ${\IM} A_0$ receives dominantly contributions from QCDP but also from EWP. The contributions from the chromo-magnetic penguins turn out to be
negligible \cite{Constantinou:2017sgv,Buras:2018evv}.
${\IM} A_2$ receives contributions exclusively from EWP. Keeping
this in mind it is useful to write
\cite{Aebischer:2020jto} 
\begin{equation}
  \left(\frac{\varepsilon'}{\varepsilon}\right)_{\text{SM}} =
 \left(\frac{\varepsilon'}{\varepsilon}\right)_{\text{QCDP}}
- \left(\frac{\varepsilon'}{\varepsilon}\right)_{\text{EWP}}
\ee
with
\be
 \left(\frac{\varepsilon'}{\varepsilon}\right)_{\text{QCDP}}=
\IM\lambda_{\rm t}\cdot 
\big(1-\hat\Omega_{\rm eff}\big) \big[15.4\,\bsi(\mu^*)-2.9 \big],
\label{QCDP}
\ee
\be
 \left(\frac{\varepsilon'}{\varepsilon}\right)_{\text{EWP}}=
\IM\lambda_{\rm t}\cdot \big[8.0\,\bei(\mu^*) -2.0\big].
\label{EWP}
\ee
This formula includes NLO QCD corrections to the QCDP contributions and NNLO contributions to EWP ones mentioned previously. The coefficients in this formula
and the parameters $\bsi$ and $\bei$, conventionally normalized to unity at the factorization scale, are scale dependent. Here
we will set  $\mu^*=1\GeV$ because at this scale it is most convenient to compare the values for $\bsi$ and $\bei$ obtained in the three  non-perturbative approaches LQCD, ChPT and DQCD that we already encountered in the context of the
$\Delta I=1/2$ rule.

The $\bsi$ and $\bei$ represent the relevant hadronic matrix elements of
the dominant QCDP and EWP operators, respectively:
\be
Q_6 = (\bar s_{\alpha} d_{\beta})_{V-A}\!\!\sum_{q=u,d,s,c,b} (\bar q_{\beta} q_{\alpha})_{V+A},
\ee
\be
Q_8 = \frac{3}{2}\,(\bar s_{\alpha} d_{\beta})_{V-A}\!\!\sum_{q=u,d,s,c,b}
      e_q\,(\bar q_{\beta} q_{\alpha})_{V+A},
      \ee
      with $V-A=\gamma_\mu(1-\gamma_5)$ and $V+A=\gamma_\mu(1+\gamma_5)$.
      They are then left-right operators with large hadronic matrix
      elements which assures their dominance over left-left operators.
      The remaining QCDP and EWP operators, represented here by $-2.9$ and
      $-2.0$, respectively, play subleading roles. Current-current operators
      $Q_{1,2}$ that played crucial role in the case of the $\Delta I=1/2$ rule
      do not contribute to $\epe$ because their WCs are real. In obtaining the formulae in
      (\ref{QCDP}) and (\ref{EWP}) it is common to use the experimental
      values for the real parts of $A_{0,2}$ in (\ref{N1}). Finally,
      $\IM\lambda_{\rm t}=\IM(V_{ts}^*V_{td})\approx 1.4\times 10^{-4}$.
      
There are two main reasons
why $Q_8$ can compete with $Q_6$ here despite the smallness of the electroweak couplings in the WC of $Q_8$ relative to the QCD one in the WC of $Q_6$.  In the basic formula (\ref{eprimea}) for $\epe$ its contribution is 
enhanced relative to the one of $Q_6$ by the  factor ${\RE A_0/\RE A_2}=22.4$. In addition its WC is enhanced for the large top-quark mass which is not the case {for} $Q_6$   \cite{Flynn:1989iu,Buchalla:1989we}.
    
 In the three non-perturbative
approaches the values of $\bsi$ and $\bei$ were found at $\mu=1\GeV$ to be:
\begin{eqnarray}\label{LATB8}
 \hspace{-8mm}  \bsi(1\GeV)   = 1.49 \pm 0.25,\qquad \bei(1\GeV) = 0.85 \pm 0.05\,,  &  (\text{RBC-UKQCD}-2020)\\  
  \label{ChPTB}
 \hspace{-8mm}     \bsi(1\GeV)   = 1.35 \pm 0.20, \qquad  \bei(1\GeV) 
    = 0.55 \pm 0.20\, , &  (\text{ChPT}-2019)\\
 \label{DQCDB}
  \hspace{-8mm}    \bsi(1\GeV)   \le 0.6,\qquad \bei(1\GeV) = 0.80 \pm 0.10\,, & (\text{DQCD}-2015)
 \end{eqnarray}
 While the large $\bsi$ and $\bei<1.0$ from LQCD has until now no  physical
 interpretation, the pattern found in  ChPT
 results apparently from  final state interactions (FSI)  that
 enhance $\bsi$ above unity and suppress 
 $\bei$ below it  \cite{Antonelli:1995gw,Bertolini:1995tp,Pallante:1999qf,Pallante:2001he}. The suppression
 of $\bsi$ and $\bei$ below unity in the DQCD approach comes
 from the meson evolution \cite{Buras:2015xba} which is required to have a proper
 matching with the WCs of QCDP and EWP operators.
The meson evolution is absent in present ChPT calculations and it is argued 
in \cite{Buras:2016fys} that including it in ChPT calculations will lower
$\bsi$ below unity. On the other hand adding non-leading FSI in {the} DQCD approach
would raise $\bsi$ above $0.6$.  Nevertheless  $\bsi\le 1.0$ is expected to be satisfied
even after the inclusion of FSI in DQCD.

 Moreover, while ChPT and DQCD use $\hat\Omega_{\rm eff}^{(8)}=(17\pm9)\, 10^{-2}$
 and $\hat\Omega_{\rm eff}^{(9)}=(29\pm7)\, 10^{-2}$, respectively, as already stated above, RBC-UKQCD still uses $\hat\Omega_{\rm eff}=0$.
These differences in the values of $\bsi$, $\bei$ and $\hat\Omega_{\rm eff}$
 imply significant differences in $\epe$ presented by these three groups:
 \be
  \label{RBCUKQCD}
  (\epe)_{\rm SM} 
  = (21.7 \pm 8.4) \times 10^{-4}, \qquad (\text{RBC-UKQCD})
  \ee
  from the RBC-UKQCD collaboration \cite{RBC:2020kdj} which uses
 $\hat\Omega_{\rm eff}=0$. Here  statistical, parametric and systematic uncertainties have been added in
quadrature. Next
  \be
  \label{Pich}
  (\epe)_\text{SM}   = (14 \pm 5) \times 10^{-4}, \qquad (\text{ChPT})
  \ee
  from ChPT \cite{Cirigliano:2019ani}.
  The large error is related to the problematic matching
  of LD and SD contributions in this approach which can be traced back
  to the absence of meson evolution in this approach. Finally
  \be\label{AJBFINAL}
 (\epe)_{\rm SM}= (5\pm2)\cdot 10^{-4}, \qquad (\text{DQCD})
 \ee
 from DQCD \cite{Buras:2015xba,Buras:2016fys,Buras:2020wyv}, where
 $\bsi\le1.0$ has been used.
 
 While the results in (\ref{RBCUKQCD}) and (\ref{Pich}) are fully consistent
 with the data {shown in (\ref{EXP}), the DQCD result in (\ref{AJBFINAL})  implies a significant anomaly and NP at work.
   Clearly, the confirmation of the DQCD result by LQCD is highly important.
   However, to this end it is desired that other LQCD collaborations get involved in these  calculations.

Let us end this presentation  with  good news. There is
a very good agreement between
LQCD and DQCD as far as EWP contribution to $\epe$ is concerned. This implies
that this contribution to $\epe$, that is unaffected by leading isospin breaking corrections, is already
known within the SM with acceptable accuracy:
\be
  \label{EWPSM}
  (\epe)^{\text{EWP}}_\text{SM}   = - (7 \pm 1) \times 10^{-4} \,,\qquad
  (\text{LQCD~and~DQCD}).
\ee
Because both LQCD and DQCD can perform much better in the case of EWP than in the case of QCDP I expect that this result will remain with us for the coming years. On the other hand,
 the value from ChPT of $\bei\approx 0.55$ \cite{Cirigliano:2019ani} implies 
using (\ref{EWP})  that the EWP contribution is roughly by a factor of 2 below the result in (\ref{EWPSM}).
Let us hope that at KM60 we will know which prediction
 is right. Further summaries  can be found in 
\cite{Buras:2020xsm,Buras:2020wyv,Buras:2022cyc} and details in original references.

Let me finish this 2023 summary of the status of $\epe$ in the SM by the
following comment. I have no doubt that the present DQCD result is significantly
closer to the true value of $\epe$ in the SM than the ones obtained presently by LQCD and ChPT because this approach includes
both the full isospin breaking effects and the meson evolution.  Both
play crucial role in the suppression of QCDP to $\epe$. But LQCD did not take
isospin breaking effects into account and ChPT did not include meson evolution.
This evolution has been demonstrated in \cite{Buras:2018lgu} to be crucial for
the  understanding of the pattern of the BSM hadronic matrix elements entering the $K^0-\bar K^0$ mixing obtained by ETM, SWME and RBC-UKQCD lattice collaborations \cite{Carrasco:2015pra,Jang:2015sla,Garron:2016mva,Boyle:2017skn,Boyle:2017ssm}. Including the SM matrix element, five matrix elements are involved and
not having the meson evolution one would miss the numerical values of some of them by factors of two.
This means that the DQCD approach passed another very  non-trivial test and
that LQCD calculations include QCD dynamics represented by  the meson evolution
although it is hidden in their extensive numerical computations.
However, it is another story that is summarized in Section 13.2.4 of my book
\cite{Buras:2020xsm} with further details in \cite{Buras:2018lgu}.
Also the fact that  LQCD calculations \cite{Bae:2014sja,Carrasco:2015pra,Boyle:2018eor} confirmed after
many years the DQCD result for the $\hat B_K$ parameter \cite{Bardeen:1987vg,Buras:2014maa} is a success for this approach. More on this  in chapter 7 of my book.

If one day the anomaly in $\epe$ will be confirmed by LQCD, it will be important
to take into account possible contributions of operators absent in the SM.
Anticipating this, the matrix elements of such new operators have been calculated in DQCD in  \cite{Aebischer:2018rrz}. Let us hope that LQCD collaborations will also calculate these matrix elements one day. Having these matrix elements
master formulae for $\epe$ beyond the SM have been derived and analysed
in detail \cite{Aebischer:2018csl,Aebischer:2018quc}. Finally,  the NLO QCD results for WCs in WET obtained in \cite{Aebischer:2021raf} allowed to derive NLO WET master formula for $\epe$ \cite{Aebischer:2021hws}.

\section{WET and SMEFT beyond Leading Order}\label{sec:SMEFT}
We have just mentioned  Weak Effective Theory (WET) and Standard Model Effective Field Theory (SMEFT) in the context of $\epe$ but it should
be emphasized that both  play these days very important roles in the tests of the SM and of the NP beyond it in the full particle physics with the pioneering
work done in \cite{Buchmuller:1985jz,Grzadkowski:2010es}.
Recent reviews can be found in \cite{Brivio:2017vri} for Higgs physics
and for flavour physics in Chapter 14 of \cite{Buras:2020xsm}. An excellent
review discussing the fundamentals of the SMEFT and various strategies
for phenomenology appeared in \cite{Isidori:2023pyp}.

The status of the short distance calculations in these theories is not yet
at the level of the ones within the SM that we summarized briefly in
Section~\ref{THF}. Most analyses these days are still performed in the LO
approximation with the corresponding RG technology of all these operators presented for SMEFT in \cite{Jenkins:2013zja,Jenkins:2013wua,Alonso:2013hga} and in
\cite{Jenkins:2017dyc,Aebischer:2017gaw}  for WET. However, also in this case, in order to increase the precision of the theory 
it is necessary to go beyond the LO analyses both in the WET and also in
the SMEFT. To this end, it is
mandatory to include first in the renormalization group (RG) analyses in these
theories the one-loop matching contributions, both between these two theories
as well as when passing thresholds at which heavy particles are integrated out.
But this is not the whole story. To complete a NLO analysis and remove
various renormalization scheme (RS) dependences in the one-loop matching also
two-loop anomalous dimensions of all operators in the WET and SMEFT have
to be included. This is a big challange because of the large number of operators
involved in both theories. Yet, during the last years significant progress
in reaching these goals has been made by various authors.

The present status of these efforts in the case of non-leptonic {meson} $\Delta F=1$ decays and $\Delta F=2$ quark mixing processes is as follows:
\begin{itemize}
\item
  The matchings in question are known by now both at tree-level \cite{Jenkins:2017jig} and one-loop
  level \cite{Dekens:2019ept}. Previous partial results can be found, for example, in \cite{Aebischer:2015fzz,Bobeth:2016llm,Bobeth:2017xry,Hurth:2019ula, Endo:2018gdn,Grzadkowski:2008mf}.
\item
  The one-loop anomalous dimension matrices (ADMs)  relevant for the RG in WET \cite{Jenkins:2017dyc,Aebischer:2017gaw} and SMEFT \cite{Jenkins:2013zja,Jenkins:2013wua,Alonso:2013hga} are
  also known.
\item
  The two-loop QCD ADMs relevant for RG evolutions for both    $\Delta F=2$ and $\Delta F=1$ non-leptonic transitions in WET are also known \cite{Buras:2000if,Aebischer:2021raf,Aebischer:2020dsw}.
\item
  The two-loop QCD ADMs relevant for RG evolutions of   $\Delta F=2$ transitions in SMEFT are also known \cite{Aebischer:2022anv} and the ones for $\Delta F=1$ transitions should be known soon.
\item
  Master formulae for $\Delta F=2$ amplitudes both in WET ans SMEFT have been
  presented in \cite{Aebischer:2020dsw} and illustrated with tree-level exchanges of heavy gauge bosons $(Z^\prime,~G^\prime)$ and corresponding heavy scalars.
\item
  On-shell methods for the computation of the one-loop and two-loop ADMs in the
  SMEFT have been developed in \cite{EliasMiro:2020tdv,Bern:2020ikv,Machado:2022ozb}. They allow a good insight into the flavour structure of the ADMs.
\item
 Very recently NLO RG analysis for scalar leptoquarks has been performed in \cite{Banik:2023ogi}.
\end{itemize}

In view of many operators involved, it was crucial to develop sophisticated computing tools
for the SMEFT. The most recent summary of the existing tools can be found in
\cite{Aebischer:2023irs}.

\section{Summary and Shopping List}
There is no question about that during last 50 years a dramatic progress has been made in the theory of Kaons. But as evident from last two chapters of my book
\cite{Buras:2020xsm}  there is still a lot to be done if we want to reach Zeptouniverse with the
help of FCNCs processes one day. Most important in this search will be in my view correlations between processes in different meson systems and also correlations with lepton flavour violation, electric dipole moments and anomalous magnetic moments. Also constraints from high energy processes have to be taken into account.
Here the SMEFT will play a crucial role as well. In principle I could now
make a list of many processes which will play crucial role in this expedition
but I think a better idea is to ask interested readers to read the last two chapters of my book, in particular Chapter 20 in which one finds the shopping list.

While, in contrast to some theorists, I am optimistic that there is NP at
scales lower than the Planck scale, the signs of no NP in $\Delta F=2$ processes
and the absence of fully convincing anomalies in flavour changing processes, could be a warning that a desert up to Planck scale except possibly for right-handed neutrinos remains a possibility. Fortunately, the numerous predictions for
various flavour observables based on the assumption of no NP in the $\Delta F=2$
Archipelago \cite{Buras:2022wpw,Buras:2022qip} 
prepare us not only for this possibility but also for demonstrating
that such a nightmare scenario is not present in nature. Here Kaon physics
will play the crucial role because as demonstrated in \cite{Buras:2014zga} rare $K$ decays are more sensitive to very high scales than it is the case of rare $B$ decays. Therefore I am looking forward to improved measurements of the
branching ratios for $\kpn$, $\klpn$, $K_{L,S}\to \mu^+\mu^-$, $K_L\to\pi^0\ell^+\ell^-$ at CERN and J-PARC and their correlations with rare $B$ decays measured
at CERN and by Belle II. In this context it is of great interest to see
whether various $B$ physics anomalies will remain when new data from Belle II
will be available. For recent updates see
\cite{Greljo:2022jac,Alguero:2023jeh,Capdevila:2023yhq}.

{\bf Acknowledgements}

I would like to thank the organizers for inviting me to give this talk at
this very important event and my numerous collaborators listed in the references
list for exciting time we spent together. I am also thankful to the unknown
referee for finding several typos in the original version and very positive
comments about my contribution.
 Financial support from the Excellence Cluster ORIGINS,
funded by the Deutsche Forschungsgemeinschaft (DFG, German Research
Foundation), Excellence Strategy, EXC-2094, 390783311 is acknowledged.

\renewcommand{\refname}{R\lowercase{eferences}}

\addcontentsline{toc}{section}{References}

\bibliographystyle{JHEP}

\bibliography{Bookallrefs}

\end{document}